\documentclass[twocolumn,showpacs,preprintnumbers,a4paper,pre,notitlepage]{revtex4-1}
\usepackage{graphicx, subfigure}
\usepackage{amssymb, amsmath,amssymb,amsfonts}
\usepackage{amsthm,mathrsfs,amsopn}
\usepackage{dcolumn}
\usepackage{bm}
\usepackage{color}
\usepackage[utf8]{inputenc}
\usepackage{mathtools}

\theoremstyle{plain} 

\begin{document}

\title{Global synchronization on time-varying higher-order structures}

\author{Md Sayeed Anwar} 
\author{Dibakar Ghosh}  
\affiliation{Physics and Applied Mathematics Unit, Indian Statistical Institute, 203 B. T. Road, Kolkata 700108, India} 	
\author{Timoteo Carletti}
\affiliation{Department of Mathematics and Namur Institute for Complex Systems, naXys, University of Namur, 2 rue Graf\'e, Namur B5000, Belgium}

\begin{abstract}
Synchronization has received a lot of attention from the scientific community for systems evolving on static networks or higher-order structures, such as hypergraphs and simplicial complexes. In many relevant real world applications, the latter are not static but do evolve in time, in this paper we thus discuss the impact of the time-varying nature of high-order structures in the emergence of global synchronization.
 To achieve this goal we extend the master stability formalism to account, in a general way, for the additional contributions arising from the time evolution of the higher-order structure supporting the dynamical systems. The theory is successfully challenged against two illustrative
examples, the Stuart-Landau nonlinear oscillator and the Lorenz chaotic oscillator.
\end{abstract}

\maketitle
\section{Introduction}
\label{sec:intro}
In the realm of complex systems, synchronization refers to the intriguing ability of coupled nonlinear oscillators to self-organize and exhibit a collective unison behavior without the need for a central controller~\cite{arenasreview}. This phenomenon, observed in a wide range of human-made and natural systems~\cite{boccaletti2018synchronization}, continues to inspire scientists seeking to unravel its underlying mechanisms.
\par To study synchronization, network science has proved to be a powerful and effective framework. Here, the interconnected nonlinear oscillators are represented as nodes, while their interactions are depicted as links~\cite{barabasibook}. However, the classical static network representation has its limitation in modeling many empirical systems, such as social networks~\cite{wasserman1994social}, brain networks~\cite{valencia2008dynamic,bassett2011dynamic}, where the connections among individual basic units are adaptable enough to be considered to evolve through time. Therefore, the framework of networks has been generalized as to include time-varying networks~\cite{holme2012temporal,masuda2016guide}, whose connections vary with time. The results presented in this framework support the claim that synchronization is enhanced by the dynamics of the supporting medium~\cite{ghosh2022synchronized, CarlettiFanelli2022,intra2}.
\par Another intrinsic limitation of networks is due to their capability to only model pairwise interactions. To go beyond this issue, scholars have brought to the fore the relevance of higher-order structures, which surpass the traditional network setting that models the interactions between individual basic units only through pairwise links~\cite{carlettifanellinicoletti2020,battiston2020networks,battiston2021physics,majhi2022dynamics,boccaletti2023structure}. By considering the simultaneous interactions of many agents, higher-order structures, namely hypergraphs~\cite{berge1973graphs} and simplicial complexes~\cite{bianconi2021higher}, offer a more comprehensive understanding of complex systems. These higher-order structures have been proven to produce novel features in various dynamical processes, including consensus~\cite{neuhauser2020multibody,schaub2}, random walks~\cite{CarlettiEtAl2020,schaub2020random}, pattern formation~\cite{carlettifanellinicoletti2020,muolo2023turing,gao2023turing}, synchronization~\cite{carlettifanellinicoletti2020,skardal2020higher,simplicialsync4,carletti2023global,anwar2022intralayer,anwar2022stability}, social contagion and epidemics~\cite{iacopini2019simplicial,chowdhary2021simplicial}. Nevertheless, the suggested framework is not sufficiently general for describing systems with many-body interactions that vary with time. As an example, group interactions in social systems have time-varying nature as the interactions among groups of individuals are not always active but rather change throughout time~\cite{cencetti2021temporal}. Some early works have begun to investigate the time-varying aspect of many-body interactions in various dynamical processes. For instance, time-varying group interactions have been demonstrated to influence the convergence period of consensus dynamics~\cite{schaub2} and to predict the onset of endemic state in epidemic spreading~\cite{chowdhary2021simplicial}.        
\par The present work is motivated by these recent research directions, and it aims to take one step further by considering the impact of time-varying higher-order structures in the synchronization of nonlinear oscillators. In this context, a preliminary effort has been reported in~\cite{anwar2022synchronization}, that investigates synchronization in time-varying simplicial complexes, limited only to fast switching~\cite{stilwell2006sufficient,petit2017theory} among distinct static simplicial configurations, implying that the  time scale of the simplicial evolution is exceedingly fast compared to that of the underlying dynamical system. In contrast, in the present work, we allow the higher-order structures to evolve freely with time, thus removing any limitations on the imposed time evolution of the higher-order structure. We present the results in the framework of hypergraphs, but they hold true also for simplicial complexes. Under such broad circumstances, we develop a theory to determine the conditions ensuring the stability of a globally synchronized state that generalizes the Master Stability Equation~\cite{msf} to a setting where the time evolution of underlying higher-order structures is explicitly considered. The generalized framework we discuss here assumes that the coupling functions cancel out when the dynamics of individual oscillators are identical, which is a necessary condition that must be met for the extended system to have a synchronous solution and it has been frequently used in the literature across various domains. The developed theory reveals that the consideration of temporality in group interactions can induce synchronization more easily than static group interactions, tested on higher-order structures of coupled Stuart Landau oscillators and paradigmatic Lorenz systems.        

\section{The model}
\label{sec:themodel}
To start with, let us consider a $m$-dimensional dynamical system whose time evolution is described by the following ordinary differential equation
\begin{equation}
\label{eq:isolated}
\frac{d\vec{x}}{dt} = \vec{f}(\vec{x})\, ,
\end{equation}
where $\vec{x}\in\mathbb{R}^{m}$ denotes the state vector and $\vec{f}:\mathbb{R}^m\rightarrow \mathbb{R}^m$ some smooth nonlinear function; let us assume moreover that system~\eqref{eq:isolated} exhibits an oscillatory behavior, being the latter periodic or irregular; we are thus considering the framework of generic nonlinear oscillators. Let us now consider $n$ identical copies of system~\eqref{eq:isolated} coupled by a symmetric higher-order structure; namely, we allow the nonlinear oscillators to interact in couples, as well as in triplets, quadruplets, and so on, up to interactions among $D+1$ units. We can thus describe the time evolution of the state vector of the $i$-th unit by
\begin{equation}
\dot{\vec{x}}_i = \vec{f}(\vec{x_i}) + \sum\limits_{d=1}^D q_d\sum\limits_{j_1,\dots,j_d=1}^n A_{ij_1\dots j_d}^{(d)}(t)\vec{g}^{(d)}(\vec{x}_i,\vec{x}_{j_1},\dots,\vec{x}_{j_d})\, , 
\label{eq:dyn_main}
\end{equation}
where for $d=1,\dots,D$, $q_d>0$ denotes the coupling strength, $\vec{g}^{(d)}:\mathbb{R}^{(d+1)m}\rightarrow \mathbb{R}^m$ the nonlinear coupling function and $\mathbf{A}^{(d)}(t)$ the tensor encoding which units are interacting together. More precisely ${A}^{(d)}_{ij_1\dots j_d}(t)=1$ if the units $i,j_1,\dots ,j_d$ do interact at time $t$, observe indeed that such tensor depends on time, namely the intensity of the coupling as well which units are coupled, do change in time. Finally, we assume the time-varying interaction to be symmetric, namely if ${A}^{(d)}_{ij_1\dots j_d}(t)=1$, then ${A}^{(d)}_{\pi(ij_1\dots j_d)}(t)=1$ for any permutation $\pi$ of the indexes $i,j_1,\dots , j_d$. Let us emphasize that we consider the number of nodes to be fixed, only the interactions change in time; one could relax this assumption by considering to have a sufficiently large reservoir of nodes, from which the core of the system can recruit new nodes or deposit unused nodes.

\par Let us fix a periodic reference solution, $\vec{s}(t)$, of system~\eqref{eq:isolated}. We are interested in determining the conditions under which the orbit $(\vec{s}(t),\dots,\vec{s}(t))^\top$ is a solution of the coupled system~\eqref{eq:dyn_main}, and moreover it is stable, namely the $n$ units globally synchronize and behave at unison. A necessary condition is that the coupling functions vanish once evaluated on such orbit, i.e., $\vec{g}^{(d)}(\vec{s},\dots,\vec{s})=0$, for $d=1,\dots, D$. This assumption is known in the literature as {\em non-invasive} condition. 

\par For the sake of pedagogy, we will hereby consider a particular case of non-invasive couplings and we will refer the interested reader to Appendix~\ref{app:noninv} for a general discussion. We are thus assuming the coupling functions $\vec{g}^{(d)}$ to be {\em diffusive-like}, namely for each $d$ there exists a function $\vec{h}^{(d)}:\mathbb{R}^{dm}\rightarrow \mathbb{R}^m$ such that
\begin{equation}
\label{eq:dyn2}
\vec{g}^{(d)}(\vec{x}_i,\vec{x}_{j_1},\dots,\vec{x}_{j_d})=\vec{h}^{(d)}(\vec{x}_{j_1},\dots,\vec{x}_{j_d})-\vec{h}^{(d)}(\vec{x}_{i},\dots,\vec{x}_{i})\, .
\end{equation}
In this way we can straightforwardly ensure that the coupling term in Eq.~\eqref{eq:dyn2} vanishes once evaluated on the orbit $(\vec{s}(t),\dots,\vec{s}(t))^\top$, allowing thus to conclude that the latter is also a solution of the coupled system. 

\par To study the stability of the reference solution, let us now perturb the synchronous solution $(\vec{s}(t),\dots,\vec{s}(t))^\top$ with a spatially inhomogeneous term, meaning that $\forall i\in\{1,\dots,n\}$ we define $\vec{x}_i=\vec{s}+\delta\vec{x}_i$. Substituting the latter into Eq.~\eqref{eq:dyn_main} and expanding up to the first order, we obtain 
\begin{widetext}
\begin{equation}
\label{eq:linearized}
\delta\dot{\vec{x}}_i  =   \frac{\partial \vec{f}}{\partial \vec{x}_i}\Big\rvert_{\vec{s}}\delta\vec{x}_i+\sum_{d=1}^D q_d\sum_{j_1,\dots,j_d=1}^n B_{ij_1\dots j_d}(t)  \sum_{\ell=1}^{d} \frac{\partial \vec{h}^{(d)}}{\partial \vec{x}_{j_\ell}}\Big\rvert_{(\vec{s},\dots,\vec{s})}\delta\vec{x}_{j_\ell}\, ,
\end{equation}
\end{widetext}
where 
\begin{eqnarray*}
B_{ij_1}(t)&=& A_{ij_1}^{(1)}(t)- k^{(1)}_i(t)\delta_{ij_1} \, ,\\
B_{ij_1j_2}(t)&=&A_{ij_1j_2}^{(2)}(t)-2k_{i}^{(2)}(t)\delta_{ij_1j_2}\, , \dots\\
B_{ij_1j_2\dots j_D}(t)&=&A_{ij_1j_2\dots j_D}^{(D)}(t)-D!k_{i}^{(D)}(t)\delta_{ij_1j_2\dots j_D}\, ,
\end{eqnarray*} 
being $\delta_{ij_1j_2\dots j_D}$ the generalized multi-indexes Kronecker-$\delta$, and the (time-varying) $d$-degree of node $i$ is given by
\begin{equation}
\label{eq:ki}
k_{i}^{(d)}(t)=\frac{1}{d!}\sum_{j_1,..,j_d=1}^n A_{ij_1\dots j_d}^{(d)}(t)\, ,
\end{equation}
which represents the number of hyperedges of order $d$ incident to node $i$ at time $t$. Observe that if $\mathbf{A}^{(d)}$ is weighted, then $k_{i}^{(d)}(t)$ counts both the number and the weight, it is thus the generalization of the strength of a node. Let us now define 
\begin{equation}
\label{eq:kij}
k_{ij}^{(d)}(t)=\frac{1}{(d-1)!}\sum_{j_1,...,j_{d-1}}^n A_{ijj_1\dots j_{d-1}}^{(d)}(t)\, ,
\end{equation}
namely the number of hyperedges of order $d$ containing both nodes $i$ and $j$ at time $t$. Again, once $\mathbf{A}^{(d)}$ is weighted, then $k_{ij}^{(d)}(t)$ generalizes the link strength. Let us observe that because of the invariance of $\mathbf{A}^{(d)}$ under index permutation, we can conclude that $k_{ij}^{(d)}(t)=k_{ji}^{(d)}(t)$. Finally, we define the generalized time-varying higher-order Laplacian matrix for the interaction of order $d$ as 
\begin{equation}
\label{eq:Lij}
L_{ij}^{(d)}(t)= 
\begin{cases} -d!k_{i}^{(d)}(t) & \text{if $i=j$} \\ (d-1)!k_{ij}^{(d)}(t) & \text{if $i\neq j$}
\end{cases}\, .
\end{equation}
Observe that such a matrix is symmetric because of the assumption of the tensors $\mathbf{A}^{(d)}$. Let us also notice the difference in sign with respect to other notations used in the literature.

\par We can then rewrite Eq.~\eqref{eq:linearized} as follows
\begin{widetext}
\begin{eqnarray}
\label{eq:linearized2}
\delta\dot{\vec{x}}_i  &=&   \frac{\partial \vec{f}}{\partial \vec{x}_i}\Big\rvert_{\vec{s}}\delta\vec{x}_i+\sum_{d=1}^D q_d\left[\sum_{j_1=1}^n \frac{\partial \vec{h}^{(d)}}{\partial \vec{x}_{j_1}}\Big\rvert_{(\vec{s},\dots,\vec{s})}\delta\vec{x}_{j_1}\sum_{j_2,\dots,j_d=1}^n B_{ij_1\dots j_d}(t)  +\dots+ \sum_{j_d=1}^n \frac{\partial \vec{h}^{(d)}}{\partial \vec{x}_{j_d}}\Big\rvert_{(\vec{s},\dots,\vec{s})}\delta\vec{x}_{j_d}\sum_{j_1,\dots,j_{d-1}=1}^n B_{ij_1\dots j_d}(t)\right]\notag\\
&=&   \frac{\partial \vec{f}}{\partial \vec{x}_i}\Big\rvert_{\vec{s}}\delta\vec{x}_i+\sum_{d=1}^D q_d\sum_{j=1}^n L^{(d)}_{ij}(t)\left[\frac{\partial \vec{h}^{(d)}}{\partial \vec{x}_{j_1}} +\dots+ \frac{\partial \vec{h}^{(d)}}{\partial \vec{x}_{j_d}}\right]_{(\vec{s},\dots,\vec{s})}\delta\vec{x}_{j}\, ,
\end{eqnarray}
\end{widetext}
where we used the fact the $\frac{\partial \vec{h}^{(d)}}{\partial \vec{x}_{j_1}} +\dots+ \frac{\partial \vec{h}^{(d)}}{\partial \vec{x}_{j_d}}$ is independent from the indexes being the latter just place holders to identify the variable with respect to the derivative has to be done. Finally, by defining  
\begin{eqnarray*}
\mathbf{J}_{f}&:=&\frac{\partial \vec{f}}{\partial \vec{x}_{i}}\Big\rvert_{\vec{s}(t)}\text{ and }\\
\mathbf{J}_{h^{(d)}}&:=&\sum_{\ell=1}^d \frac{\partial \vec{h}^{(d)}}{\partial \vec{x}_{j_\ell}}\Big\rvert_{(\vec{s}(t),\dots,\vec{s}(t))}\forall d\in\{1,\dots,D\}\, , 
\end{eqnarray*} 
we can rewrite Eq.~\eqref{eq:linearized2} in compact form
\begin{equation}
\label{eq:linearized3}
\delta\dot{\vec{x}}_i  =   \mathbf{J}_{f}\delta\vec{x}_i+\sum_{d=1}^D q_d\sum_{j=1}^n L^{(d)}_{ij}(t)\mathbf{J}_{h^{(d)}}\delta\vec{x}_{j}\, .
\end{equation}
This is a non-autonomous linear differential equation determining the stability of the perturbation $\delta\vec{x}_i$, for instance, by computing the largest Lyapunov exponent. To make some analytical progress in the study of Eq.~\eqref{eq:linearized3}, we will consider two main directions: the functions $\vec{h}^{(d)}$ satisfy the condition of {\em natural coupling} (see Section~\ref{ssec:natcoup}) or the higher-order structures exhibit {\em regular topologies} (see Section~\ref{ssec:regstr}). The aim of each assumption is to disentangle the dependence of the nonlinear coupling functions from the higher-order Laplace matrices and thus achieve a better understanding of the problem under study.

\subsection{Natural coupling}
\label{ssec:natcoup}
Let us assume the functions $\vec{h}^{(d)}$ to satisfy the condition of natural coupling, namely
\begin{equation}
\label{eq:natcoupl}
\vec{h}^{(D)}(\vec{x},\dots,\vec{x})=\dots=\vec{h}^{(2)}(\vec{x},\vec{x})=\vec{h}^{(1)}(\vec{x})\, ,
\end{equation}
that implies $\mathbf{J}_{h^{(1)}}=\mathbf{J}_{h^{(2)}}=\dots=\mathbf{J}_{h^{(D)}}$
and it allows to eventually rewrite Eq.~\eqref{eq:linearized3} as follows
\begin{equation}
\label{eq:linearized4}
\delta\dot{\vec{x}}_i  =   \mathbf{J}_{f}\delta\vec{x}_i+\sum_{j=1}^n M_{ij}(t)\mathbf{J}_{h^{(1)}}\delta\vec{x}_{j}\, ,
\end{equation}
where
\begin{equation}
\label{eq:defM}
M_{ij}(t)  :=  \sum_{d=1}^D q_d L^{(d)}_{ij}(t)\quad \forall i,j=1,\dots n\, .
\end{equation}

\par Let us observe that the matrix $\mathbf{M}(t)$ is a Laplace matrix; it is non-positive definite (as each one of the $\mathbf{L}^{(d)}(t)$ matrices does for any $d=1,\dots, D$ and any $t>0$, and $q_d>0$), it admits $\mu^{(1)}=0$ as eigenvalue associated to the eigenvector $\phi^{(1)}=(1,\dots,1)^\top$ and it is symmetric. So there exists an orthonormal time-varying eigenbasis, $\phi^{(\alpha)}(t)$, $\alpha=1,\dots,n$, for $\mathbf{M}(t)$ with associated eigenvalues $\mu^{(\alpha)} \leq 0$. Let us define~\cite{CarlettiFanelli2022} the $n\times n$ time dependent matrix $\mathbf{c}(t)$ that quantifies the projections of the time derivatives of the eigenvectors onto the independent eigendirections, namely
\begin{equation}
\label{eq:cab}
\frac{d \vec{\phi}^{(\alpha)}(t)}{dt}=\sum_{\beta}c_{\alpha\beta}(t)\vec{\phi}^{(\beta)}(t)\quad\forall \alpha=1,\dots, n\, .
\end{equation}
By recalling the orthonormality condition
\begin{equation*}
\left(\vec{\phi}^{(\alpha)}(t)\right)^\top\cdot \vec{\phi}^{(\beta)}(t)=\delta_{\alpha \beta}\, ,    
\end{equation*}
we can straightforwardly conclude that $\mathbf{c}$ is a real skew-symmetric matrix with a null first row and first column, i.e., $c_{\alpha\beta}+c_{\beta\alpha}=0$ and $c_{1\alpha}=0$.

\par To make one step further, we consider Eq.~\eqref{eq:linearized4}, and we project it onto the eigendirections, namely we introduce $\delta\vec{x}_i=\sum_\alpha \delta\hat{\vec{x}}_{\alpha}\phi^{(\alpha)}_i$ and recalling the definition of $\mathbf{c}$ we obtain
\begin{equation}
\label{eq:GLHGlinalpha3}
\frac{d\delta\hat{\vec{x}}_{\beta}}{dt} = \sum_\alpha c_{\beta\alpha}(t)\delta\hat{\vec{x}}_{\alpha}+\left[\mathbf{J}_{f}+ \mu^{(\beta)}(t)\mathbf{J}_{h^{(1)}}\right]\delta\hat{\vec{x}}_{\beta}\, . 
\end{equation}
Let us observe that the latter formula and the following analysis differ from the one presented in~\cite{vangorder2} where the perturbation is assumed to align onto a single mode, a hypothesis that ultimately translates in the stationary of the Laplace eigenvectors that is $\mathbf{c}=\mathbf{0}$. The same assumption is also at the root of the results by~\cite{ZS2021}; indeed, commuting time-varying networks implies to deal with a constant eigenbasis. In conclusion, Eq.~\eqref{eq:GLHGlinalpha3} returns the more general description for the projection of the linearized dynamics on a generic time-varying Laplace eigenbasis, and thus allowing us to draw general conclusions without unnecessary simplifying assumptions.  

\subsection{Regular topologies}
\label{ssec:regstr}
An alternative approach to study Eq.~\eqref{eq:linearized3} is to assume regular topologies~\cite{muolo2023turing}, namely hypergraphs such that $\mathbf{L}^{(d)}(t) = \alpha_d \mathbf{L}^{(1)}(t)$, for $d=1,\dots,D$, with $\alpha_1=1$ and $\alpha_d\in\mathbb{R}_+$. Indeed we can use this assumption to obtain from Eq.~\eqref{eq:linearized3}
\begin{equation}
\label{eq:linearized4b}
\delta\dot{\vec{x}}_i  =   \mathbf{J}_{f}\delta\vec{x}_i+\sum_{j=1}^n L^{(1)}_{ij}(t)\mathbf{J}_{\hat{h}}\delta\vec{x}_{j}\, ,
\end{equation}
where
\begin{equation}
\label{eq:defHhhat}
\mathbf{J}_{\hat{h}}  :=  \sum_{d=1}^D q_d \alpha_d \mathbf{J}_{h^{(d)}}\, ,
\end{equation}
that results in a sort of weighted nonlinear coupling term. We can now make use of the existence of a time-varying orthonormal basis of $\mathbf{L}^{(1)}(t)$, namely $\psi^{(\alpha)}(t)$, $\alpha=2,\dots,n$, associated to eigenvalues $\Lambda^{(\alpha)} <0$, $\psi^{(1)}(t)=(1,\dots,1)^\top$ and $\Lambda^{(1)}=0$, to project $\delta\vec{x}_i$ onto the $n$ eigendirections, $\delta\vec{x}_i=\sum_\alpha \delta\tilde{\vec{x}}_{\alpha}\psi^{(\alpha)}_i$. Because the latter vary in time we need to define a second $n\times n$ time dependent matrix $\mathbf{b}(t)$ given by
\begin{equation}
\label{eq:bab}
\frac{d \vec{\psi}^{(\alpha)}(t)}{dt}=\sum_{\beta}b_{\alpha\beta}(t)\vec{\psi}^{(\beta)}(t)\quad\forall \alpha=1,\dots, n\, ,
\end{equation}
that it is again real, skew-symmetric, with a null first row and first column, i.e., $b_{\alpha\beta}+b_{\beta\alpha}=0$ and $b_{1\alpha}=0$, because of the orthonormality condition of eigenvectors. By projecting Eq.~\eqref{eq:linearized4b} onto $\psi^{(\alpha)}(t)$,  we get
\begin{equation}
\label{eq:GLHGlinalpha3bis}
\frac{d\delta\tilde{\vec{x}}_{\beta}}{dt} = \sum_\alpha b_{\beta\alpha}(t)\delta\tilde{\vec{x}}_{\alpha}+\left[\mathbf{J}_{f}+ \Lambda^{(\beta)}(t)\mathbf{J}_{\hat{h}}\right]\delta\tilde{\vec{x}}_{\beta}\, .
\end{equation}
Let us conclude by observing that the latter equation has the same structure of~\eqref{eq:GLHGlinalpha3}. Those equations determine the generalization of the Master Stability Equation to the case of time-varying higher-order structures. The time variation signature of the topology is captured by the matrices $\mathbf{c}(t)$ or $\mathbf{b}(t)$ and the eigenvectors $\mu^{(\alpha)}(t)$ or $\Lambda^{(\alpha)}(t)$, while the dynamics (resp. the coupling) in the Jacobian $\mathbf{J}_f$ (resp. $\mathbf{J}_{h^{(1)}}$ or $\mathbf{J}_{\hat{h}}$). 
\par It is important to notice that as the eigenvalues $\mu^{(1)}=0$, $\Lambda^{(1)}=0$ and the skew-symmetric matrices $\mathbf{c}(t), \mathbf{b}(t)$ have null first row and column, in analogy with the MSF approaches carried over static networks~\cite{msf} and higher-order structures~\cite{gambuzza2021stability}, also in the case of time-varying higher-order structures, we can decouple the Master Stability Equation into two components. One component describes the movement along the synchronous manifold, while the other component represents the evolution of different modes that are transverse to the synchronous manifold. The Maximum Lyapunov Exponent (MLE) associated with the transverse modes measures the exponential growth rate of a tiny perturbation in the transverse subspace. It serves as an enhanced form of Master Stability Function (MSF) and provides valuable insights into the stability of the reference orbit. For the synchronous orbit to be stable, the MLE associated to all transverse modes must be negative. Moreover, the MSF approaches applied to static networks and higher-order structures can be simplified by examining the evolution of the perturbation along each independent eigendirection associated with distinct eigenvalues of the Laplacian matrix. Let us observe that this is not possible in the present because the matrices $\mathbf{c}(t)$ and $\mathbf{b}(t)$ mix the different modes and introduce a complex interdependence among them, making it challenging to disentangle their individual contributions. For this reason, one has to address numerically the problem~\cite{CarlettiFanelli2022}.
\par To demonstrate the above introduced theory and emphasize the outcomes arising from the modified Master Stability Equations~\eqref{eq:GLHGlinalpha3} and \eqref{eq:GLHGlinalpha3bis}, we will present two key examples in the following sections. Indeed, we will utilize the Stuart-Landau limit cycle oscillator and the chaotic Lorenz system as prototype dynamical systems anchored to each individual nodes. To simplify the calculations, we assume that the hypergraph consists of only three nodes, three links and one triangle (face), whose weights change in time. Additionally, the eigenvector projection matrices $\mathbf{c}(t)$ and $\mathbf{b}(t)$ do not vary in time; this assumption results from a suitable choice of the Laplace eigenbasis as explained later in Appendix~\ref{sec:hypergraph}. Finally, to simplify the analysis we also assume the Laplace eigenvalues to be constant in time. Let us stress that despite such assumptions, the proposed framework is very general and can be applied to any time varying hypergraphs.

\section{Synchronization of Stuart-Landau oscillators coupled via time-varying higher-order networks}
\label{sec:SL}  
The aim of this section is to present an application of the theory above introduced. We decided to use the Stuart-Landau (SL) model as a prototype example for two reasons; first, it provides the normal form for a generic system close to a supercritical Hopf-bifurcation, second, because of its structure, the Jacobian of the reaction part becomes constant once evaluated on the reference orbit and this simplifies the presentation of the results.
\par A SL oscillator can be described by a complex amplitude $w$ that evolves in time according to $\dot{w}=\sigma w-\beta |w|^2w$, where $\sigma=\sigma_\Re+i\sigma_\Im$ and $\beta=\beta_\Re+i\beta_\Im$ are complex model parameters. The system admits a limit cycle solution $w_{LC}(t)=\sqrt{\sigma_\Re/\beta_\Re}e^{i\omega t}$, where $\omega=\sigma_\Im-\beta_\Im \sigma_\Re/\beta_\Re$, that is stable provided $\sigma_\Re>0$ and $\beta_\Re>0$, conditions that we hereby assume. 
\par  To  proceed in the analysis, we couple together $n$ identical SL oscillators, each described by a complex amplitude $w_j$, with $j=1,...,n$, anchored to the nodes of a time-varying hypergraph as prescribed in the previous section, namely
\begin{widetext}
\begin{equation}
\label{eq:maineqSL}
\frac{dw_j}{dt}= \sigma w_j-\beta w_j|w_j|^2 + \sum\limits_{d=1}^D q_d\sum\limits_{j_1,\dots,j_d=1}^n A_{jj_1\dots j_d}^{(d)}(t)\vec{g}^{(d)}(w_j,w_{j_1},\dots,w_{j_d})\,  .
\end{equation}
\end{widetext}
For the sake of simplicity, we restrict our analysis to pairwise and three-body interactions, namely $D=2$ in Eq.~\eqref{eq:maineqSL}. We hereby present and discuss the SL synchronization under the diffusive-like coupling hypothesis and by using two different assumptions: regular topology and natural coupling. The case of non-invasive coupling will be presented in Appendix~\ref{ssec:noninvass}.

\subsection{Diffusive-like and regular topology}
\label{ssec:difflkregtop}
Let us thus assume the existence of two functions $h^{(1)}(w)$ and $h^{(2)}(w_1,w_2)$ such that $g^{(1)}$ and $g^{(2)}$ do satisfy the diffusive-like assumption, namely
\begin{equation*}
\begin{array}{l}	
g^{(1)}(w_j,w_{j_1}) = h^{(1)}(w_{j_1})-h^{(1)}(w_j)\text{ and }\\\\g^{(2)}(w_j,w_{j_1},w_{j_2}) = h^{(2)}(w_{j_1},w_{j_2})-h^{(2)}(w_j,w_j)\, .
\end{array}
\end{equation*}
For the sake of definitiveness, let us fix
\begin{equation}
\label{eq:h1h2}
h^{(1)}(w)=w \text{ and } h^{(2)}(w_1,w_2)=w_1w_2\, ,
\end{equation}
let us observe that the latter functions do not satisfy the condition for natural coupling, indeed $h^{(1)}(w)=w\neq  w^2=h^{(2)}(w,w)$.

\par Let us assume to deal with regular topology, namely $\mathbf{L}^{(2)}=\alpha_2\mathbf{L}^{(1)}$. Hence following Eq.~\eqref{eq:defHhhat} we can define $\mathbf{J}_{\hat{h}}  =  q_1 \mathbf{J}_{h^{(1)}}+q_2 \alpha_2 \mathbf{J}_{h^{(2)}}$. Let us perturb the limit cycle solution $w_{LC}(t)=\sqrt{\sigma_\Re/\beta_\Re}e^{i\omega t}$ by defining $w_j=W_{LC}(1+\rho_j)e^{i\theta_j}$, where $\rho_j$ and $\theta_j$ are real and small functions for all $j$. A straightforward computation allows to write the time evolution of $\rho_j$ and $\theta_j$
\begin{widetext}
\begin{equation}
\label{eq:lindifflk}
 \dfrac{d}{dt}\left(\begin{matrix} {\rho_j} \\{\theta_j}\end{matrix}\right) = 
\left(\begin{matrix}
 -2\sigma_\Re & 0\\-2\beta_\Im \frac{\sigma_\Re}{\beta_\Re} & 0
\end{matrix}\right)\left(\begin{matrix} {\rho_j} \\{\theta_j}\end{matrix}\right)+\sum_\ell L_{j\ell}^{(1)} \biggl[\left(\begin{matrix}
  q_{1,\Re} & - q_{1,\Im}\\ q_{1,\Im} &   q_{1,\Re}
\end{matrix}\right)+ 2\alpha_2 \sqrt{\frac{\sigma_\Re}{\beta_\Re}}
\left(\begin{matrix}
\cos (\omega t) & - \sin (\omega t)\\ \sin (\omega t) & \cos (\omega t)
\end{matrix}\right)\left(\begin{matrix}
  q_{2,\Re} & - q_{2,\Im}\\ q_{2,\Im} &   q_{2,\Re}
\end{matrix}\right)\biggr]\left(\begin{matrix} {\rho_\ell} \\{\theta_\ell}\end{matrix}\right) \, ,
\end{equation}
\end{widetext}
where $\omega =\sigma_\Im-\beta_\Im \sigma_\Re/\beta_\Re$ is the frequency of the limit cycle solution.

By exploiting the eigenvectors $\psi^{(\alpha)}(t)$ and eigenvalues $\Lambda^{(\alpha)}(t)$ of $\mathbf{L}^{(1)}(t)$ to project the perturbation $\rho_j$ and $\theta_j$ we obtain:
\begin{widetext}
\begin{equation}
\label{eq:lindifflkproj}
 \dfrac{d}{dt}\left(\begin{matrix} {\rho_\beta} \\{\theta_\beta}\end{matrix}\right) = \sum_\alpha b_{\beta\alpha}\left(\begin{matrix} {\rho_\alpha} \\{\theta_\alpha}\end{matrix}\right)+\Bigl\{
\left(\begin{matrix}
 -2\sigma_\Re & 0\\-2\beta_\Im \frac{\sigma_\Re}{\beta_\Re} & 0
\end{matrix}\right) +\Lambda^{(\beta)} \left[\left(\begin{matrix}
  q_{1,\Re} & - q_{1,\Im}\\ q_{1,\Im} &   q_{1,\Re}
\end{matrix}\right) \\ + 2\alpha_2 \sqrt{\frac{\sigma_\Re}{\beta_\Re}}
\left(\begin{matrix}
\cos (\omega t) & - \sin (\omega t)\\ \sin (\omega t) & \cos (\omega t)
\end{matrix}\right)\left(\begin{matrix}
  q_{2,\Re} & - q_{2,\Im}\\ q_{2,\Im} &   q_{2,\Re}
\end{matrix}\right)\right]\Bigr\}\left(\begin{matrix} {\rho_\beta} \\{\theta_\beta}\end{matrix}\right) \, ,
\end{equation}
\end{widetext}
where  the matrix $\mathbf{b}$ has been defined in Eq.~\eqref{eq:bab}.

\par For the sake of definiteness and to focus on the impact of the time-varying topology, we hereby consider a simple higher-order network structure composed of $n=3$ nodes, three links and one triangle. Moreover, the eigenvalues are assumed to be constant and the time-derivative of the associated eigenvectors projected on the eigenbasis to return a constant matrix $\mathbf{b}$, for a given $\Omega\geq 0$
\begin{equation}
\label{eq:bmatrixOmega}
		\mathbf{b}=\begin{pmatrix}
			0 &0 &0 \\
			0 & 0 & \Omega \\
			0 & -\Omega & 0			
		\end{pmatrix}\, .
\end{equation} 
One can show (see Appendix~\ref{sec:hypergraph} and~\cite{CarlettiFanelli2022}) that those assumptions on the hypergraph correspond to two eigenvectors rotating in a plane orthogonal to the constant eigenvector $\psi^{(1)}\sim (1,\dots,1)^\top$ with frequency $\Omega>0$. The case $\Omega=0$ corresponds thus to a static higher-order network structure. 
\begin{figure}[ht!]
	      \includegraphics[scale=0.35]{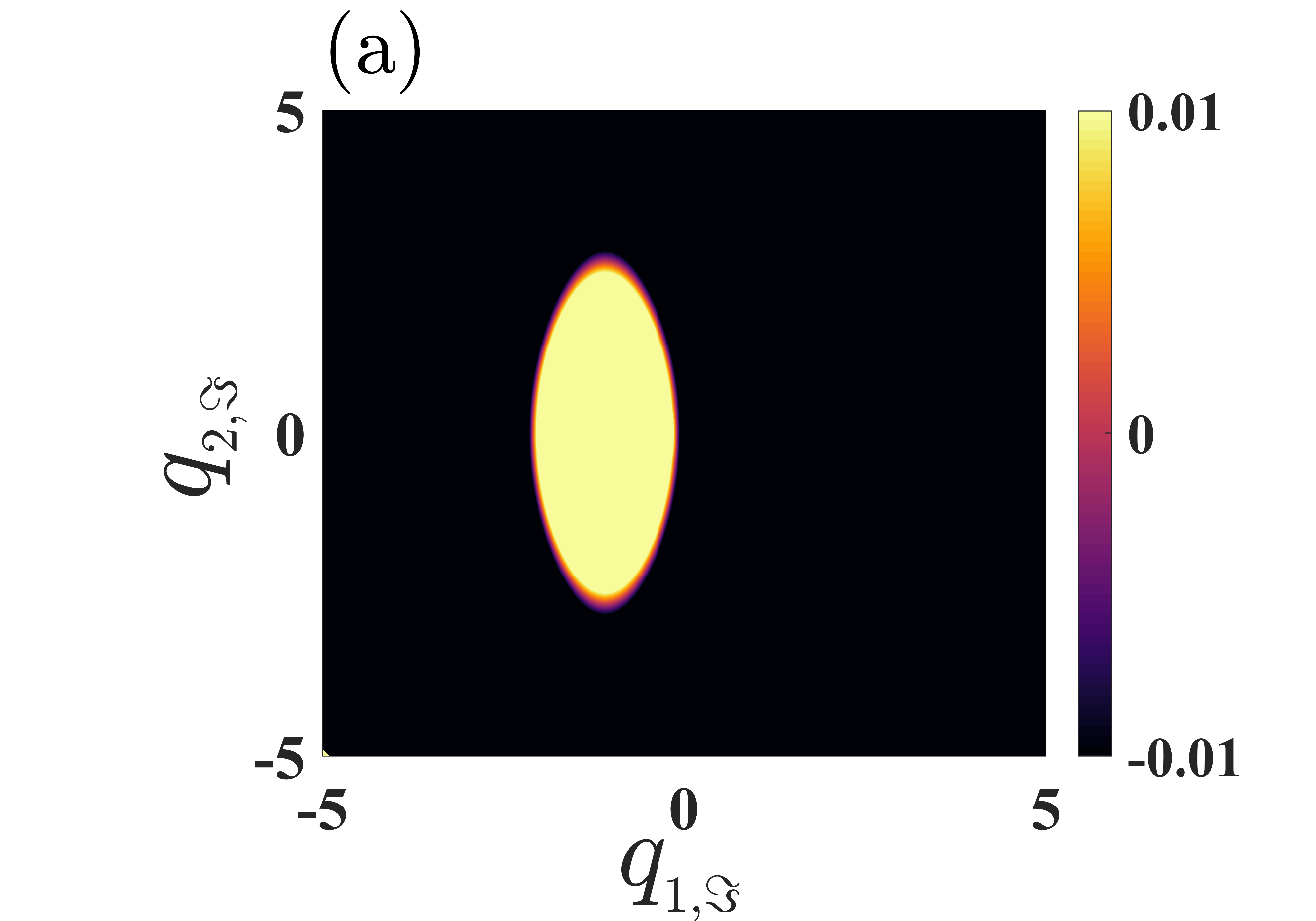}\\
	    \includegraphics[scale=0.35]{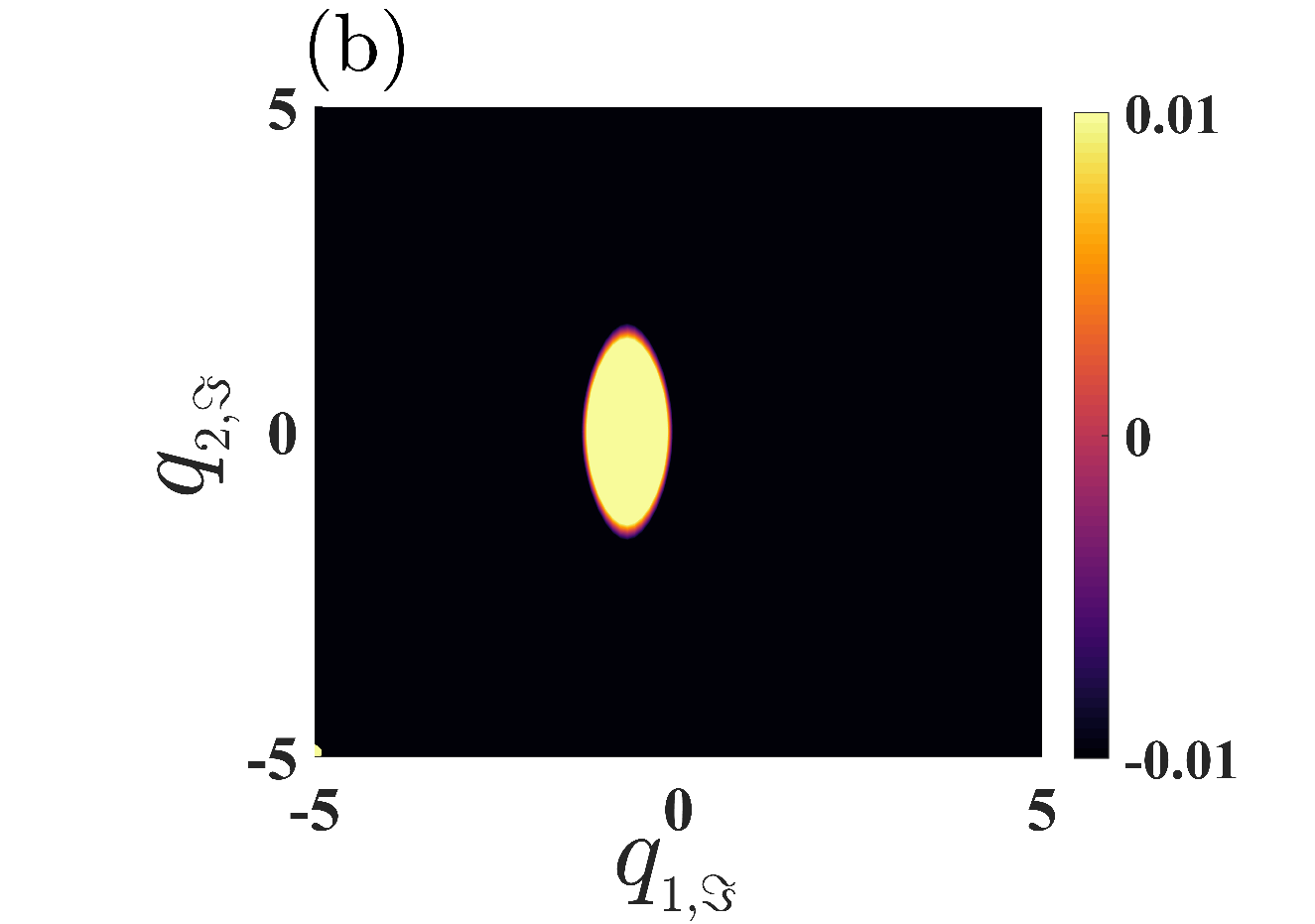}
	\caption{{\bf Synchronization on time-varying regular higher-order network of coupled SL oscillators}. We report the MSF as a function of $q_{1,\Im}$ and $q_{2,\Im}$ for two different values of $\Omega$, $\Omega=0$ (panel (a)) and $\Omega=2$ (panel (b)), by using a color code, we determine the region of stability (black) and the region of instability (yellow). The remaining parameters have been fixed at the values $\alpha_{2}=2$, $\sigma=1.0+4.3i$, $\beta=1.0+1.1i$, $q_{1,\Re}=0.1$, $q_{2,\Re}=0.1$, $\Lambda^{(2)}=-1$, and $\Lambda^{(3)}=-2$.}
	\label{SL_regular_sig1c_sig2c}
\end{figure}
\par Under those assumptions, Eq.~\eqref{eq:lindifflkproj} determines a time periodic linear system whose stability can be determined by using Floquet theory. In order to illustrate our results, we let $q_{1,\Im}$ and $q_{2,\Im}$ to freely vary in the range $[-5,5]$, while keeping fixed to generic values the remaining parameters, and we compute the Floquet eigenvalue with the largest real part, corresponding thus to the Master Stability Function (MSF) of Eq.~\eqref{eq:lindifflkproj}, as a function of $q_{1,\Im}$ and $q_{2,\Im}$. The corresponding results are shown in Fig.~\ref{SL_regular_sig1c_sig2c} for $\Omega=0$ (panel (a)) and $\Omega = 2$ (panel (b)). By a direct inspection, one can clearly conclude that the parameters region associated with a negative MSF (black region), i.e., to the stability of the SL limit cycle and thus to global synchronization, is larger for $\Omega >0$ than for $\Omega=0$.
\begin{figure}[ht!]
	\includegraphics[scale=0.35]{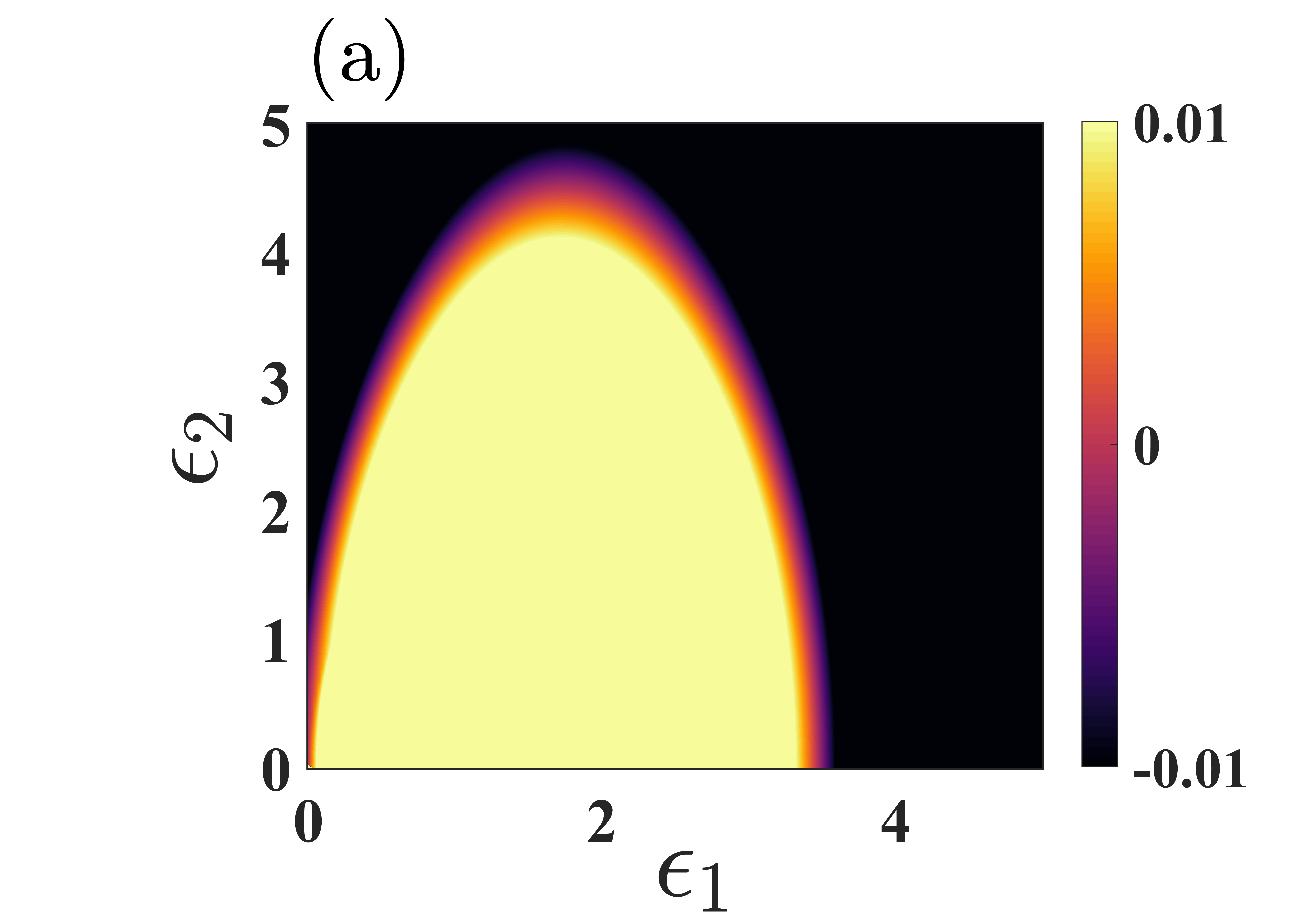}\\
	    \includegraphics[scale=0.35]{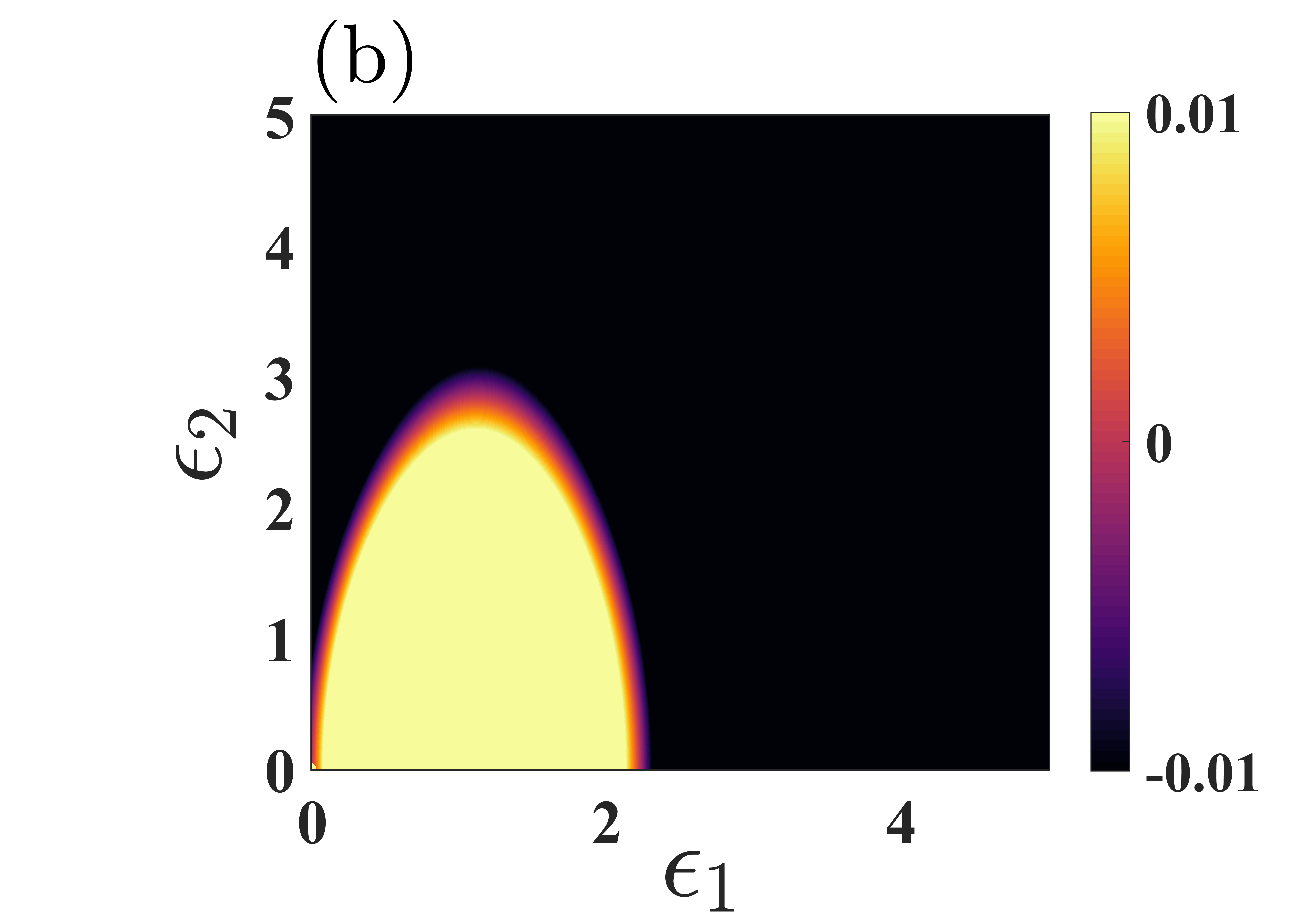}
	\caption{{\bf Synchronization on time-varying regular higher-order network of coupled SL oscillators}. The MSF is reported as a function of $\epsilon_{1}$ and $\epsilon_{2}$ for two different values of $\Omega$, $\Omega=0$ (panel (a)) and $\Omega=2$ (panel (b)). The color code represents the values of the MSF, negative values (black) while positive values (yellow). The remaining parameters have been fixed at the values $\alpha_{2}=2$, $\sigma=1.0+4.3i$, $\beta=1.0+1.1i$, $q_{1,0}=0.1-0.5i$, $q_{2,0}=0.1+0.5i$, $\Lambda^{(2)}=-1$, and $\Lambda^{(3)}=-2$.}
	\label{SL_regular_eps1_eps2}
\end{figure}
\par To study the combined effect of both coupling strengths $q_{1}$ and $q_{2}$, we set $q_{1}=\epsilon_{1}q_{1,0}$ and $q_{2}=\epsilon_{2}q_{2,0}$, and we compute the MSF as a function of $\epsilon_{1}$ and $\epsilon_{2}$, having fixed without loss of generality $q_{1,0}=0.1-0.5i$ and $q_{2,0}=0.1-0.5i$. The corresponding results are presented in Fig.~\ref{SL_regular_eps1_eps2} for static ($\Omega=0$, panel (a)) and time-varying ($\Omega=2$, panel (b)) higher-order structure. We can again conclude that the region of parameters corresponding to global synchronization (black region) is larger in the case of time-varying hypergraph than in the static case.
\begin{figure}[ht!]
	\includegraphics[scale=0.35]{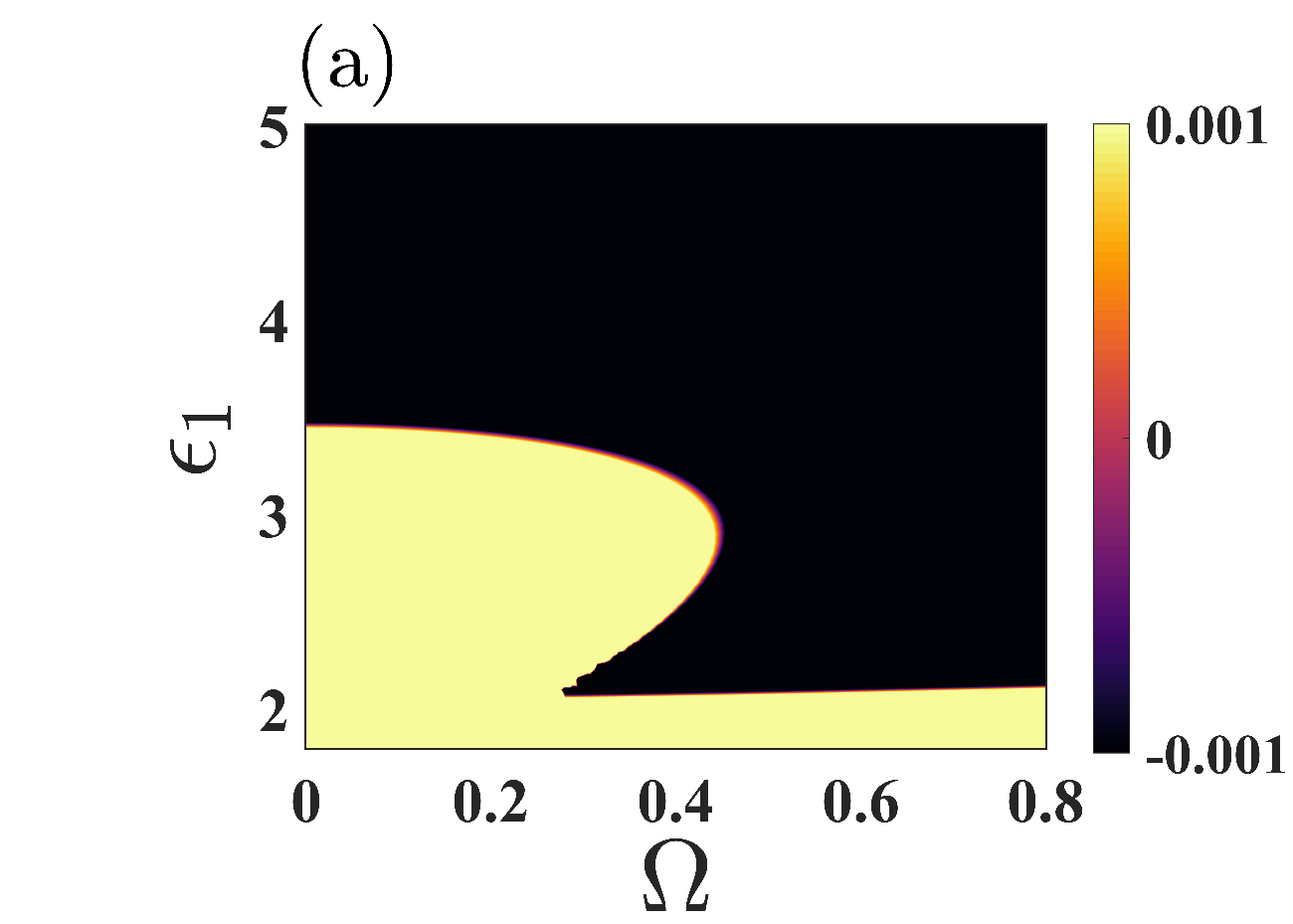}\\
	    \includegraphics[scale=0.35]{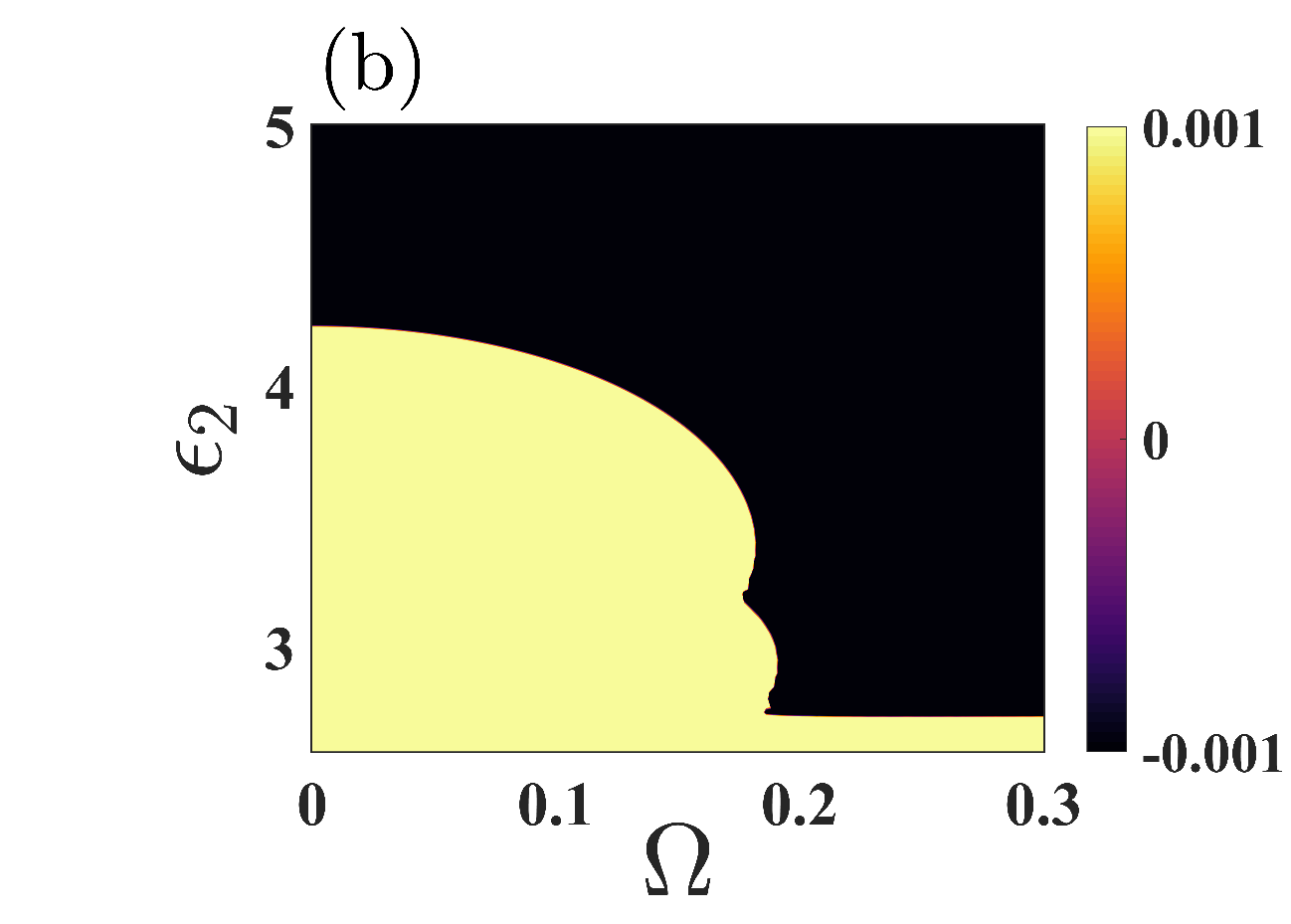}
	\caption{{\bf Synchronization domains}. We show the MSF in the plane $(\Omega,\epsilon_{1})$ (panel (a)) for $\epsilon_{2}=0.02$ and in the plane $(\Omega,\epsilon_{2})$ (panel (b)) for $\epsilon_{1}=0.02$. We can observe that in both panels, the critical value of coupling strengths $\hat{\epsilon}_j(\Omega)$ to achieve synchronization is smaller for $\Omega>0$ than for $\Omega=0$. Furthermore, in panel (a) existence of an interval $\mathcal{I}_{1}=[\Omega_{1},\Omega_{2}]$ can be observed such that for all $\Omega \in \mathcal{I}_{1}$, there exist three different values of critical coupling $\hat{\epsilon}_{1}$ for the occurrence of synchronization. In panel (b), we can observe the existence of two intervals $\mathcal{I}_{2}=[\Omega_{3},\Omega_{4}]$ and $\mathcal{I}_{3}=[\Omega_{5},\Omega_{6}]$ such that for all $\Omega \in \mathcal{I}_{2}$ there exist two critical values of $\hat{\epsilon}_{2}$ and for all $\Omega \in \mathcal{I}_{3}$ there exist three critical values of $\hat{\epsilon}_{2}$ for the emergence of synchronization. The remaining parameters are kept fixed at the values $\alpha_{2}=2$, $\sigma=1.0+4.3i$, $\beta=1.0+1.1i$, $q_{1,0}=0.1-0.5i$, $q_{2,0}=0.1+0.5i$, $\Lambda^{(2)}=-1$, and $\Lambda^{(3)}=-2$.}
	\label{SL_regular_eps_omega}
\end{figure}
\par Our last analysis concerns the relation between the frequency $\Omega$ and the size of the coupling parameters $\epsilon_{1}$, $\epsilon_{2}$, still assuming $q_{1}=\epsilon_{1}q_{1,0}$ and $q_{2}=\epsilon_{2}q_{2,0}$, on the onset of synchronization. In Fig.~\ref{SL_regular_eps_omega} we report the MSF in the plane $(\Omega,\epsilon_{1})$ for a fixed value of $\epsilon_2$ (panel (a)), and in the plane $(\Omega,\epsilon_{2})$ for a fixed value of $\epsilon_1$ (panel (b)). Let us observe that the synchronization can be easier achieved the smaller the value $\epsilon_{j}$, $j=1,2$, for which the MSF is negative, having fixed $\Omega$. Let us thus define $\hat{\epsilon}_1(\Omega)=\min \{\epsilon >0 : \mathrm{MSF}(\epsilon,\epsilon_2,\Omega)<0\}$, for fixed $\epsilon_2$, and similarly $\hat{\epsilon}_2(\Omega)$. The results of Fig.~\ref{SL_regular_eps_omega} clearly show that $\hat{\epsilon}_1(\Omega)<\hat{\epsilon}_1(0)\sim 3.5$ and $\hat{\epsilon}_2(\Omega)<\hat{\epsilon}_2(0)\sim 4.2$ and thus support our claim that time-varying structures allow to achieve synchronization easier.
\par
\begin{figure}[ht!]
		\includegraphics[scale=0.25]{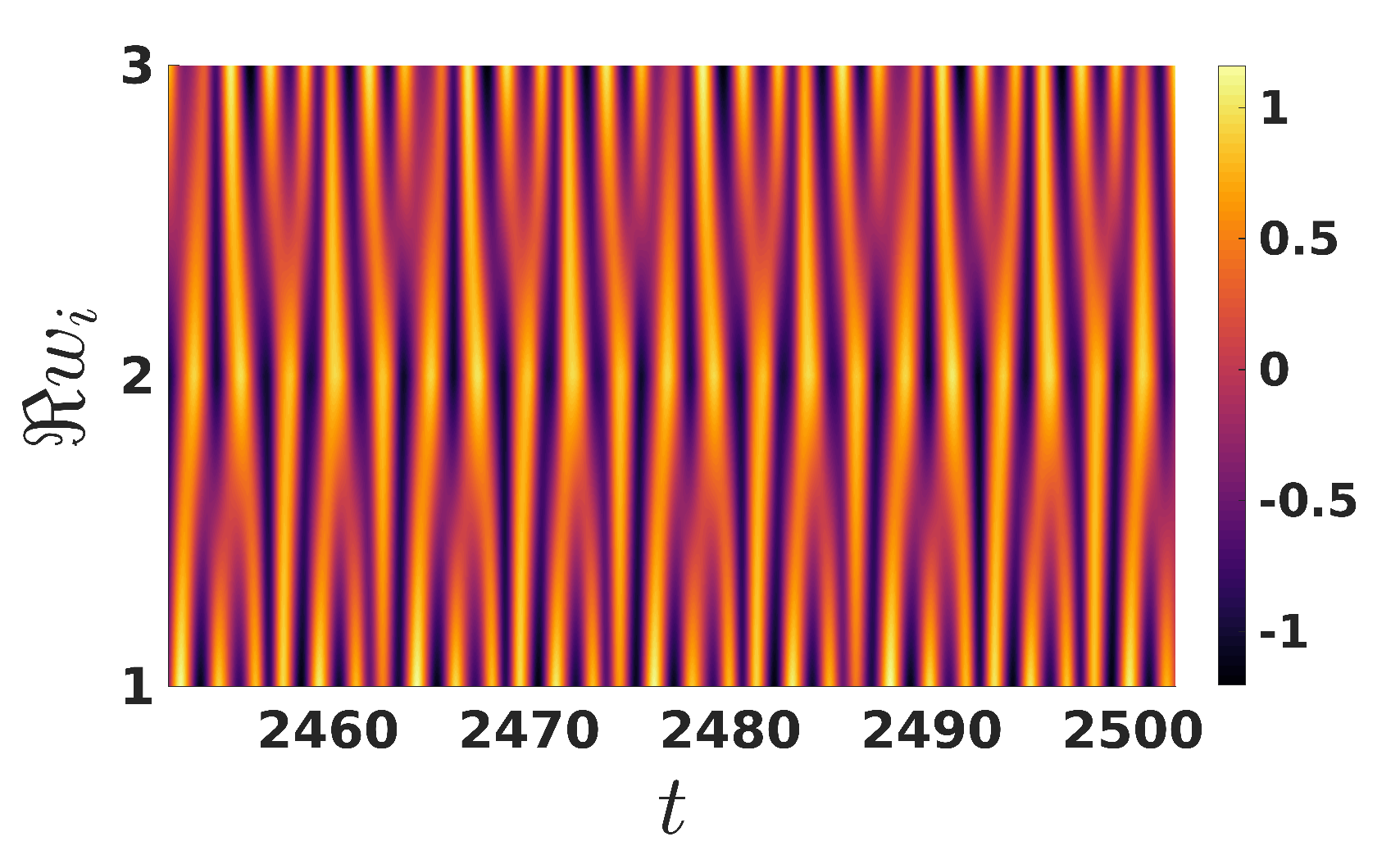}\\
	    \includegraphics[scale=0.25]{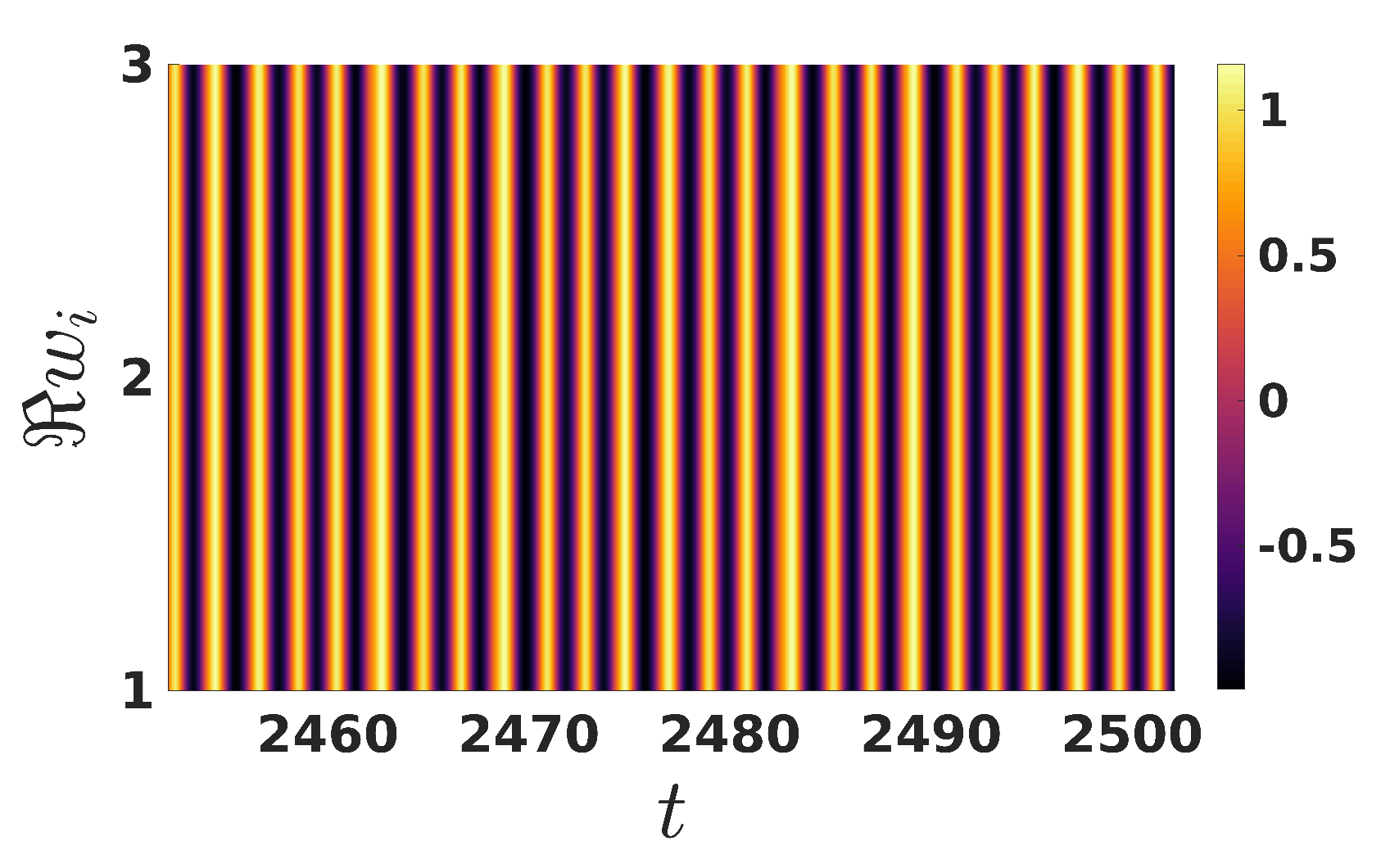}
	\caption{{\bf Temporal evolution of $\Re w_{i}$ for a suitable choice of coupling parameters $\epsilon_{1}=2.5$ and $\epsilon_{2}=0.5$}. In the top panel, we set $\Omega=0$ while $\Omega=2$ in the bottom panel. The other parameters values are the same as in Fig.~\ref{SL_regular_eps1_eps2}, i.e., $\alpha_{2}=2$, $\sigma=1.0+4.3i$, $\beta=1.0+1.1i$, $q_{1,0}=0.1-0.5i$, $q_{2,0}=0.1+0.5i$, $\Lambda^{(2)}=-1$, and $\Lambda^{(3)}=-2$.} 
	\label{time_series}
\end{figure}

 To support our analysis, we performed numerical simulations of the SL defined on the simple $3$ nodes time-varying hypergraph. We selected $(\epsilon_{1},\epsilon_{2})=(2.5,0.5)$ and the remaining parameters values as in Fig.~\ref{SL_regular_eps1_eps2}. By observing the latter figure, we conclude that for the chosen parameters, the MSF is positive if $\Omega=0$ and negative if $\Omega=2$, hence the SL should globally synchronize on the time-varying hypergraph while it would not achieve this state in the static case. Results of Fig.~\ref{time_series} confirm these conclusions; indeed, we can observe that (real part of) the complex state variable is in phase for all $i$ in the case $\Omega=2$ (right panel), while this is not clearly the case for $\Omega=0$ (left panel).

\subsection{Diffusive-like and natural coupling}
\label{ssec:difflknatcoup}

The aim of this section is to replace the condition of regular topology with a condition of natural coupling and consider thus again, a diffusive-like coupling. Let us thus consider now two functions $h^{(1)}(w)$ and $h^{(2)}(w_1,w_2)$ satisfying the natural coupling assumption, namely
\begin{equation*}
h^{(1)}(w)=h^{(2)}(w,w)\, .
\end{equation*}
For the sake of definitiveness, let us fix
\begin{equation}
\label{eq:h1h2natc}
h^{(1)}(w)=w^3 \text{ and } h^{(2)}(w_1,w_2)=w_1(w_2)^2\, .
\end{equation}

\par Consider again to perturb the limit cycle solution $w_{LC}(t)=\sqrt{\sigma_\Re/\beta_\Re}e^{i\omega t}$ by defining $w_j=W_{LC}(1+\rho_j)e^{i\theta_j}$, where $\rho_j$ and $\theta_j$ are real and small functions for all $j$. A straightforward computation allows us to write the time evolution of $\rho_j$ and $\theta_j$ as, 
\begin{widetext}
\begin{equation}
\label{eq:lindifflknatc}
 \dfrac{d}{dt}\left(\begin{matrix} {\rho_j} \\{\theta_j}\end{matrix}\right) = 
\left(\begin{matrix}
 -2\sigma_\Re & 0\\-2\beta_\Im \frac{\sigma_\Re}{\beta_\Re} & 0
\end{matrix}\right)\left(\begin{matrix} {\rho_j} \\{\theta_j}\end{matrix}\right)
+3\frac{\sigma_\Re}{\beta_\Re}\sum_\ell  M_{j\ell} \left(\begin{matrix}
\cos (2\omega t) & - \sin (2\omega t)\\ \sin (2\omega t) & \cos (2\omega t)
\end{matrix}\right)\left(\begin{matrix} {\rho_l} \\{\theta_l}\end{matrix}\right)\, ,
\end{equation}
\end{widetext}
where $\omega =\sigma_\Im-\beta_\Im \sigma_\Re/\beta_\Re$ is the frequency of the limit cycle solution and $\mathbf{M}$ is the matrix $q_1 \mathbf{L}^{(1)}(t)+q_2 \mathbf{L}^{(2)}(t)$ (see Eq.~\eqref{eq:defM}). Let us observe that in this case, the coupling parameters $q_1$ and $q_2$ should be real numbers if we want to deal with real Laplace matrices, hypothesis that we hereby assume to hold true.

\par By invoking the eigenvectors $\phi^{(\alpha)}(t)$ and eigenvalues $\mu^{(\alpha)}(t)$ of $\mathbf{M}(t)$, and the matrix $\mathbf{c}$ (see Eq.~\eqref{eq:cab}), we can project the perturbation $\rho_j$ and $\theta_j$ on the eigenbasis and thus rewrite the time variation of the perturbation as follows
\begin{widetext}
\begin{equation}
\label{eq:lindifflkprojnatc}
 \dfrac{d}{dt}\left(\begin{matrix} {\rho_\beta} \\{\theta_\beta}\end{matrix}\right) = \sum_\alpha c_{\beta\alpha}\left(\begin{matrix} {\rho_\alpha} \\{\theta_\alpha}\end{matrix}\right)+\biggl[
\left(\begin{matrix}
 -2\sigma_\Re & 0\\-2\beta_\Im \frac{\sigma_\Re}{\beta_\Re} & 0
\end{matrix}\right) +3\frac{\sigma_\Re}{\beta_\Re}  \mu^{(\beta)} \left(\begin{matrix}
\cos (2\omega t) & - \sin (2\omega t)\\ \sin (2\omega t) & \cos (2\omega t)
\end{matrix}\right)\biggr]\left(\begin{matrix} {\rho_\beta} \\{\theta_\beta}\end{matrix}\right) \, .
\end{equation}
\end{widetext}

\begin{figure}[ht!]
	\includegraphics[scale=0.35]{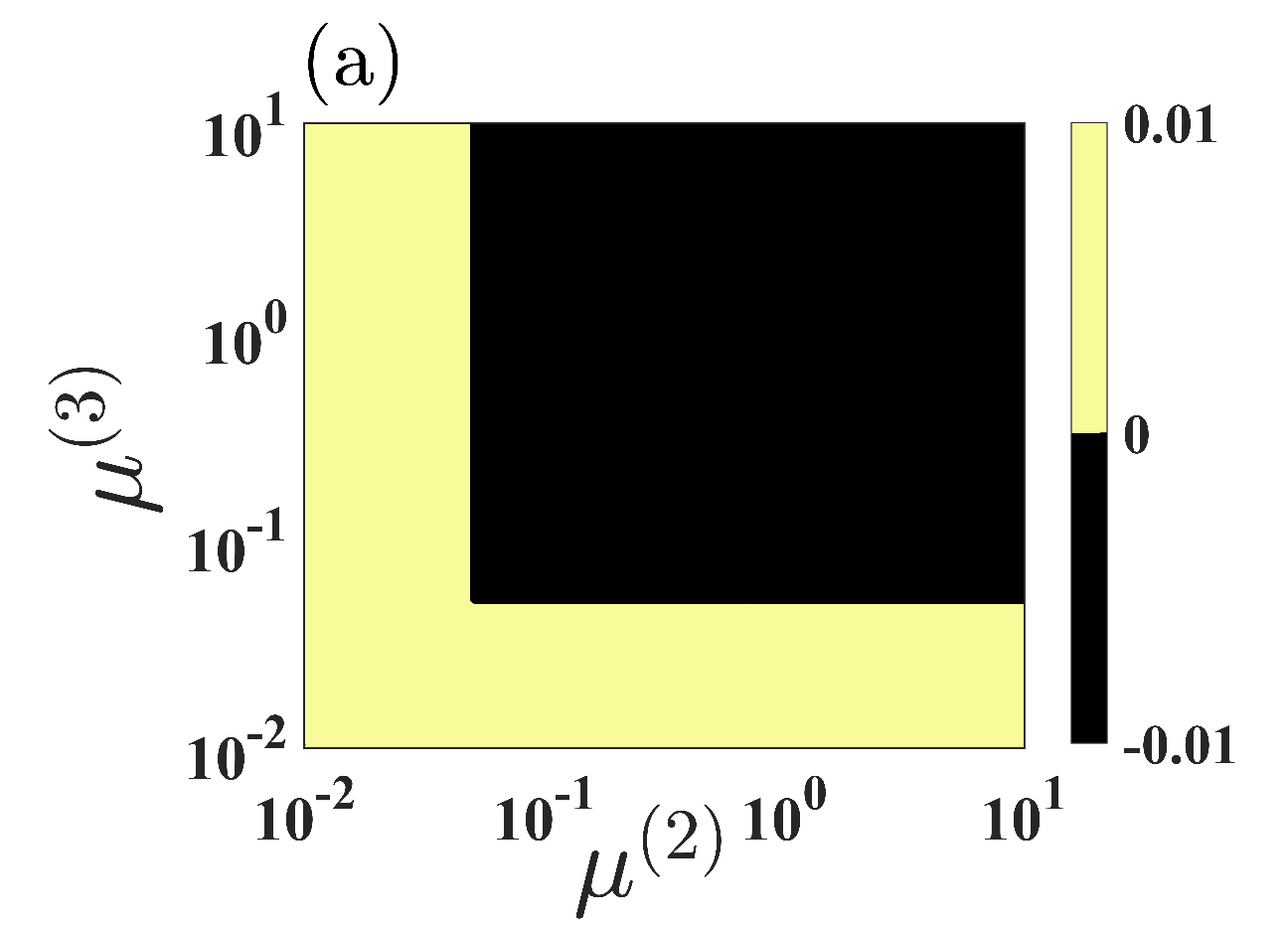}\\
	    \includegraphics[scale=0.35]{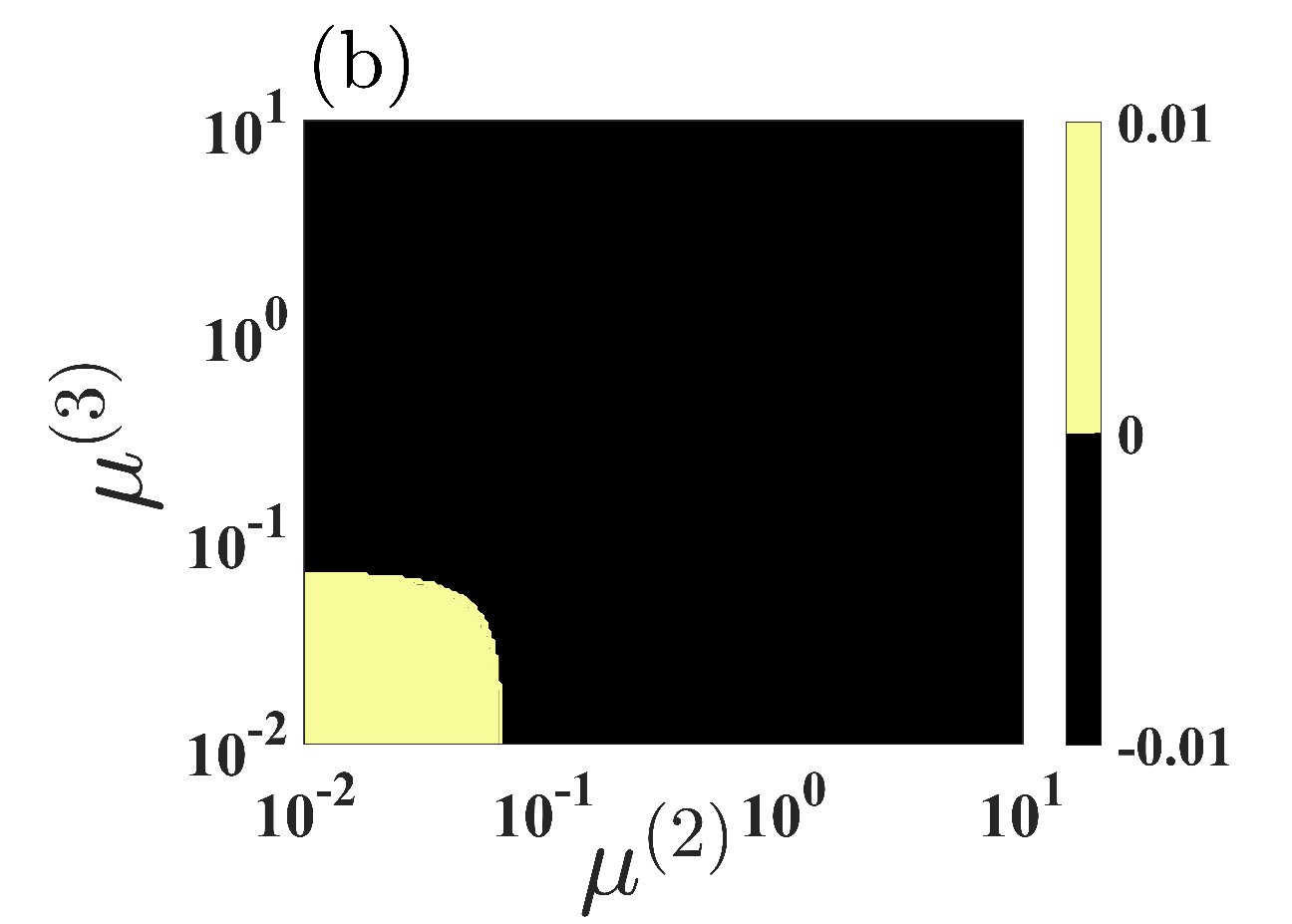}
	\caption{{\bf Synchronization on time-varying higher-order network of coupled SL oscillators with diffusive-like natural coupling}. We report the MSF as a function of the eigenvalues $\mu^{(2)}$ and $\mu^{(3)}$ for two different choices of $\Omega$, $\Omega=0$ (panel (a)) and $\Omega=2$ (panel (b)) by using a color code, black is associated to negative values while positive ones are shown in yellow. We characterize the range of the axes by considering the absolute values of the eigenvalues. The remaining parameters are kept fixed at $\sigma=1.0+4.3i$, $\beta=1.0+1.1i$.}
	\label{SL_natural_ev2_ev3}
\end{figure}
\par Let us assume again to deal with an hypergraph made by $3$ nodes and consider a time-independent matrix $\mathbf{c}$
\begin{equation*}
		\mathbf{c}=\begin{pmatrix}
			0 &0 &0 \\
			0 & 0 & \Omega \\
			0 & -\Omega & 0			
		\end{pmatrix}\, ,
\end{equation*} 
for some $\Omega \geq 0$. The eigenvalue $\mu^{(1)}=0$ of $\mathbf{M}$ determines the dynamics parallel to the synchronous manifold. On the other hand, the equations obtained for $\mu^{(2)}$ and $\mu^{(3)}$ give the dynamics of transverse modes to the synchronization manifold. Hence the MSF can be obtained by solving the latter equations and provide the conditions for a global stable synchronous solution to exist. In Fig.~\ref{SL_natural_ev2_ev3}, we show the level sets of the MSF as a function of the eigenvalues $\mu^{(2)}$ and $\mu^{(3)}$ while keeping the remaining parameters in Eq.~\eqref{eq:lindifflkprojnatc} fixed at generic nominal values. In panel (a), we consider a static hypergraph, i.e., $\Omega=0$, while in panel (b) a time-varying hypergraph, i.e., $\Omega=2$, negative values of MSF are reported in black and they correspond thus to a global synchronous state, positive values of MSF are shown in yellow; one can clearly appreciate that in the case of time-varying hypergraph, the MSF is negative for a much larger set of eigenvalues $\mu^{(2)}$ and $\mu^{(3)}$ and thus the SL system can easier synchronize.

\section{Synchronization of Lorenz systems nonlinearly coupled via time-varying higher-order networks}
\label{sec:lorenz}  
The aim of this section is to show that our results hold true beyond the example of the dynamical system shown above, i.e., the Stuart-Landau. We thus decide to present an application of synchronization for chaotic systems on a time-varying higher-order network. For the sake of definitiveness, we used the paradigmatic chaotic Lorenz model for the evolution of individual nonlinear oscillators.

\par We consider again the scenario of regular topology with the toy model hypergraph structure composed of $n=3$ nodes described previously, the whole system can thus be described by
\begin{equation}
\label{lorenz_eq}
	\begin{cases}
	   \dot{x}_{i}&=a_{1}(y_{i}-x_{i})+\epsilon_{2}\sum\limits_{j=1}^{N}\sum\limits_{k=1}^{N}A^{(2)}_{ijk}(x_{j}^{2}x_{k}-x_{i}^{3})\\
	   \dot{y}_{i}&=x_{i}(a_{3}-z_{i})-y_{i}+\epsilon_{1}\sum\limits_{j=1}^{N}A^{(1)}_{ij}(y_{j}-y_{i})\\
	   \dot{z}_{i}&=x_{i}y_{i}-a_{2}z_{i}	
	\end{cases}\, ,
\end{equation} 
where the system parameters are kept fixed at $a_{1}=10$, $a_{2}=\frac{8}{3}$, $a_{3}=28$ for which individual nodes exhibits chaotic trajectory. The pairwise and higher-order structures are related to each other by $\mathbf{L}^{(2)}=\alpha_{2}\mathbf{L}^{(1)}$. We assume the eigenvalues of the Laplacian $\mathbf{L}^{(1)}$ to be constant and the matrix $\mathbf{b}$ to be given by
\begin{equation*}
		\mathbf{b}=\begin{pmatrix}
			0 &0 &0 \\
			0 & 0 & \Omega \\
			0 & -\Omega & 0			
		\end{pmatrix}\quad \text{for some $\Omega \geq 0$.}
\end{equation*}
\par
\begin{figure}
	\includegraphics[scale=0.35]{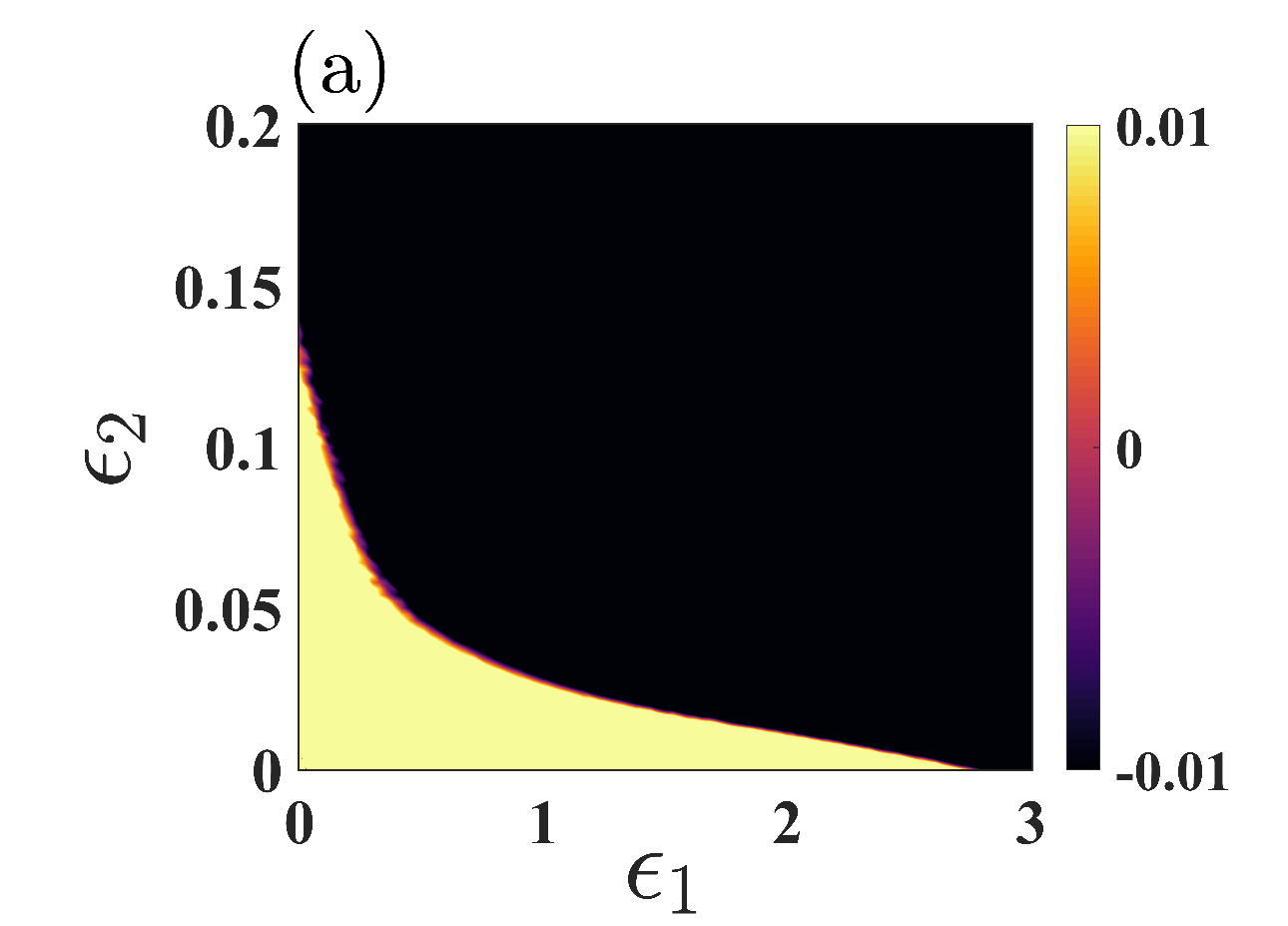}\\
	    \includegraphics[scale=0.35]{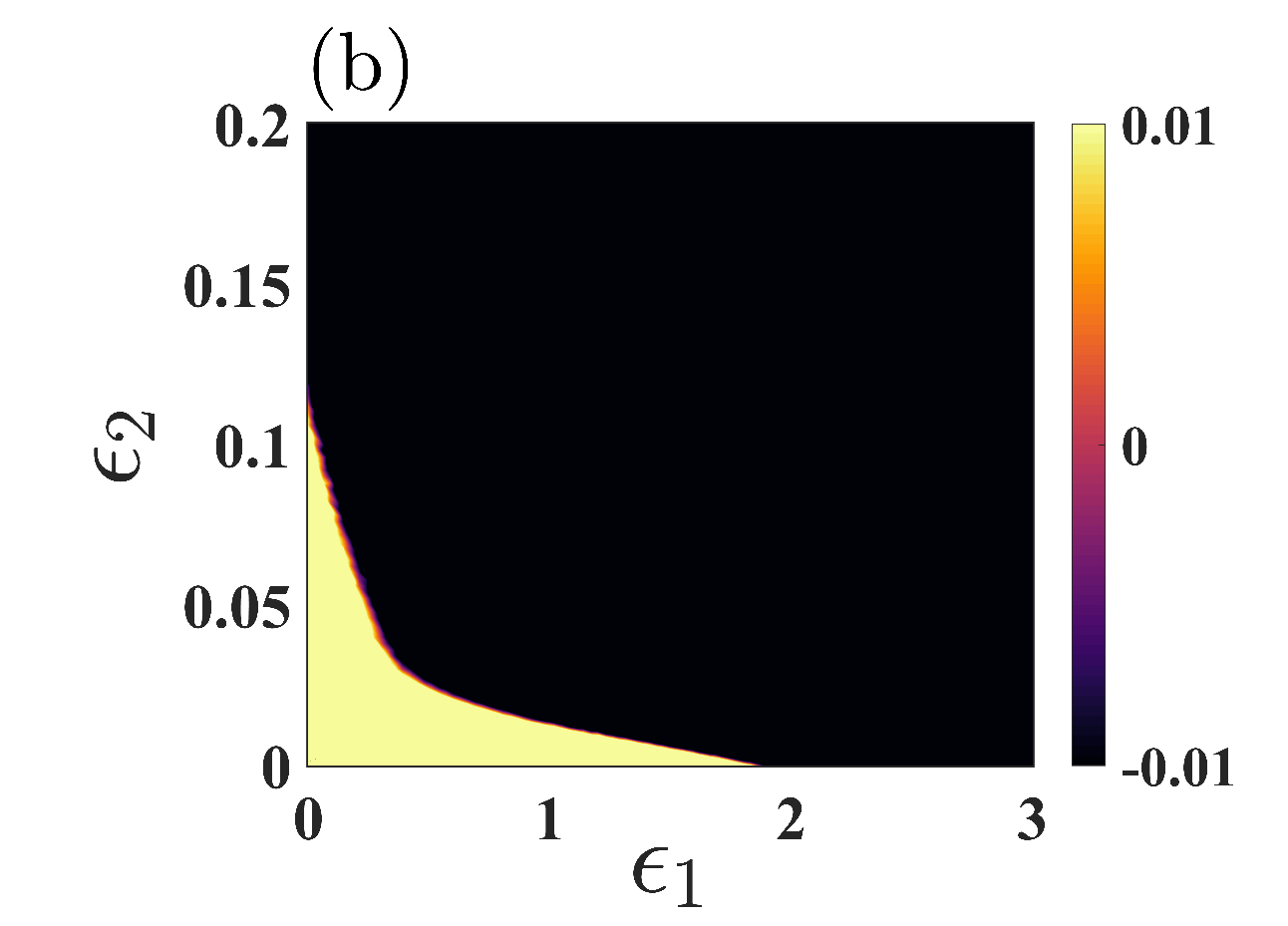}
	\caption{{\bf Synchronization on time-varying regular higher-order network of coupled Lorenz oscillators}. We report the MSF as a function of the coupling strengths, $\epsilon_{1}$ and $\epsilon_{2}$, for two different values of $\Omega$, $\Omega=0$ (panel (a)) and $\Omega=3$ (panel (b)), by using a color code, where black dots stand for a negative MSF, i.e., global synchronization, while yellow dots for a positive MSF. The remaining parameters are kept fixed at $a_{1}=10$, $a_{2}=\frac{8}{3}$, $a_{3}=28$, and $\alpha_{2}=2$.}
	\label{lorenz_regular_eps1_eps2}
\end{figure}

 Let us thus select as reference solution $\vec{s}(t)$ a chaotic orbit of the isolated Lorenz model and consider as done previously the time evolution of a perturbation about such trajectory. Computations similar to those reported above, allow to obtain a linear non-autonomous system ruling the evolution of the perturbation, whose stability can be numerically inferred by computing the largest Lyapunov exponent, i.e., the MSF. We first considered the impact of the coupling strength, $\epsilon_1$ and $\epsilon_2$ on synchronization; results are reported in Fig.~\ref{lorenz_regular_eps1_eps2} where we present the level sets of the MSF as a function of the above parameters by using a color code: black dots refer to negative MSF while yellow dots to positive MSF. The panel (a), refers to a static hypergraph, i.e., $\Omega=0$, while the panel (b) to a time-varying one, i.e., $\Omega=3$, one can thus appreciate that the latter setting allows a negative MSF for a larger range of parameters $\epsilon_1$ and $\epsilon_2$ and hence we can conclude that time-varying hypergraph enhance synchronization also in the case of chaotic oscillators.

\par
\begin{figure}
	\includegraphics[scale=0.35]{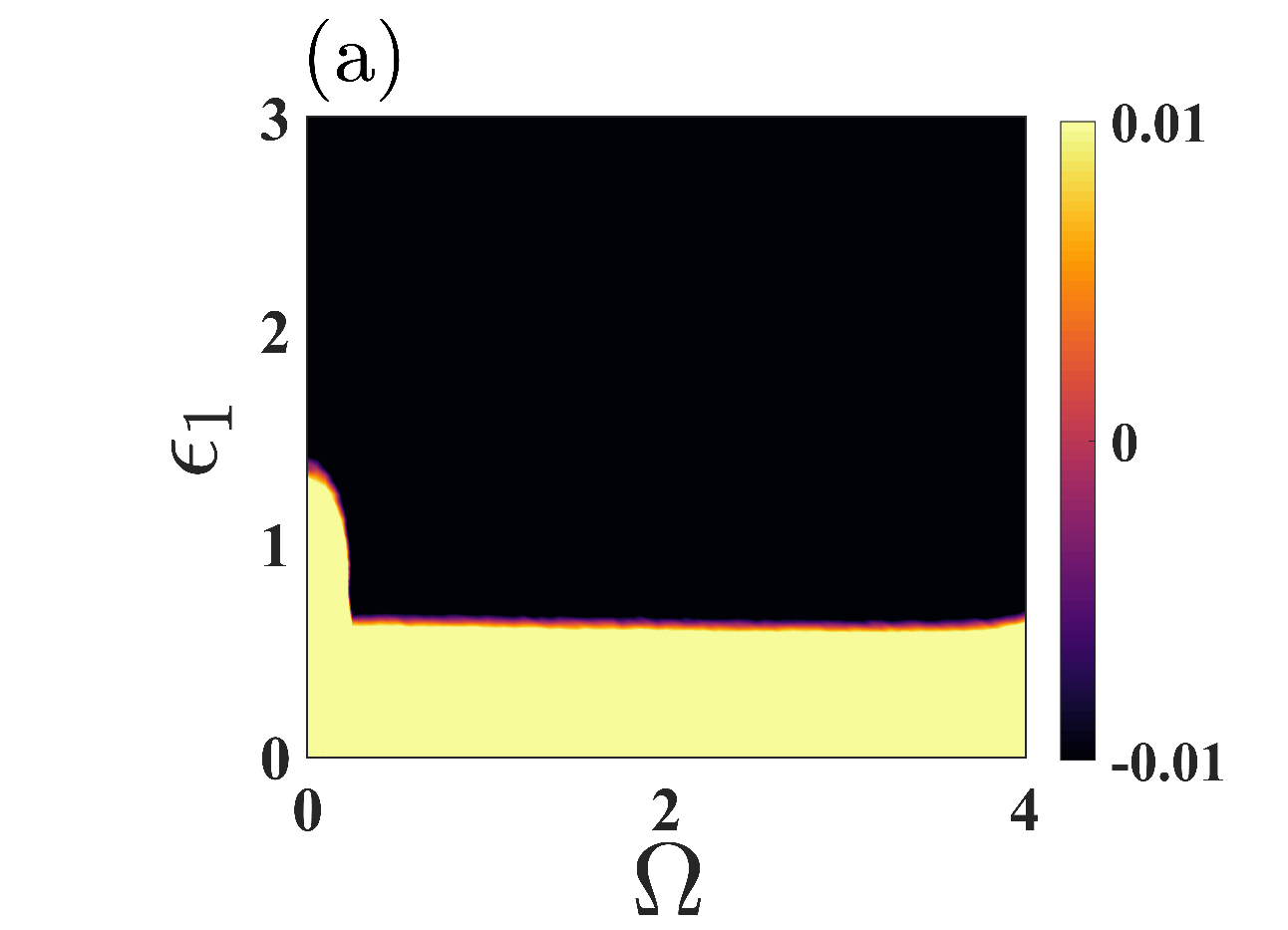}\\
	    \includegraphics[scale=0.35]{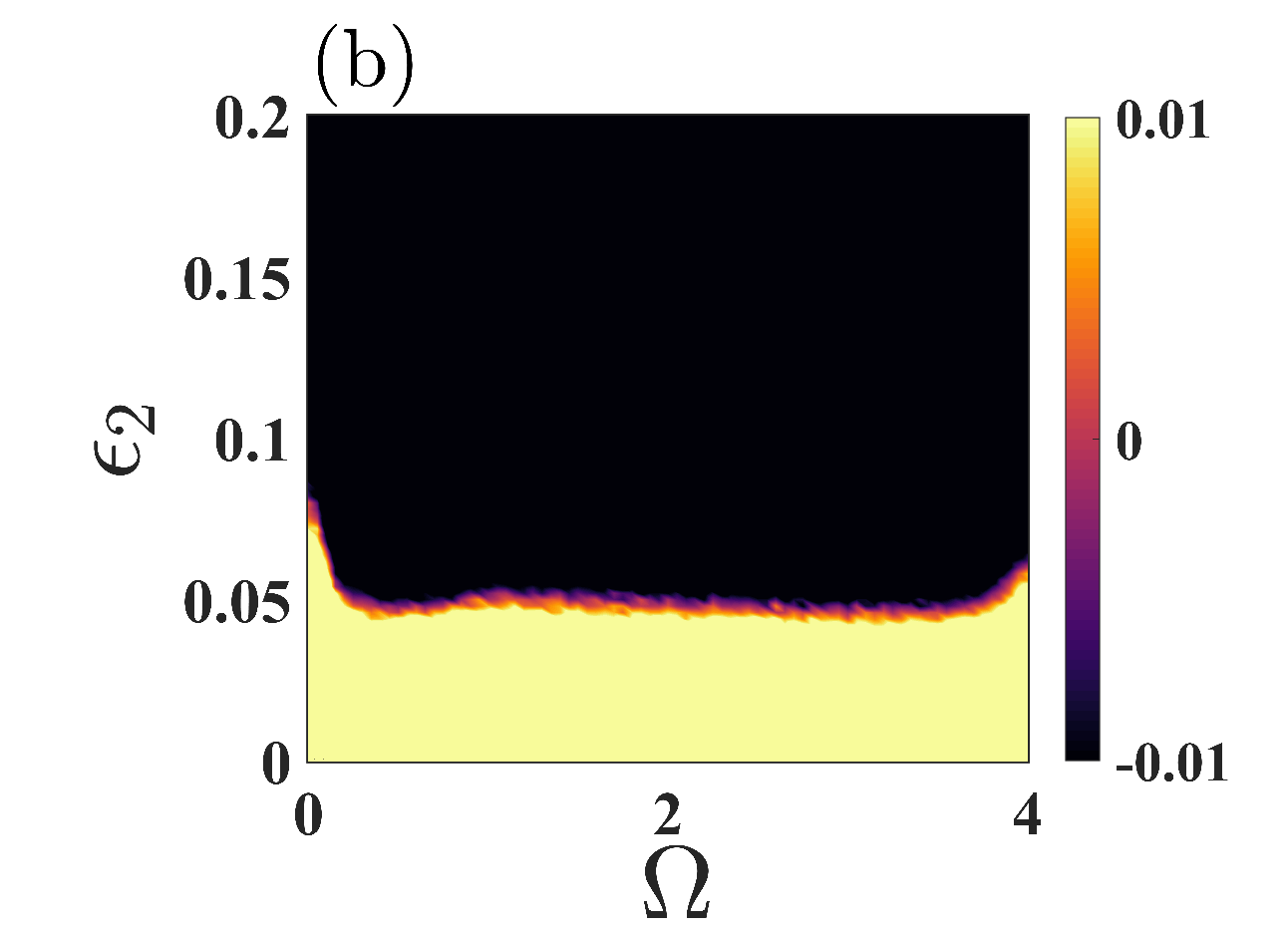}
	\caption{We show the MSF in the plane $(\Omega,\epsilon_{1})$ (panel (a)) for $\epsilon_{2}=0.01$) and in the plane $(\Omega,\epsilon_{2})$ (panel (b)) for $\epsilon_{1}=0.2$). We can observe that in both panels the critical value of coupling strengths $\hat{\epsilon}_j(\Omega)$ to achieve synchronization is smaller for $\Omega>0$ than for $\Omega=0$. This implies that synchronization can occur more easily on a time-varying higher-order structure than on a static one.}
	\label{lorenz_regular_Omega_epsi}
\end{figure}

 We conclude this analysis by studying again the relation between the frequency $\Omega$ and the size of the coupling parameters $\epsilon_{1}$, $\epsilon_{2}$ on the onset of synchronization. In Fig.~\ref{lorenz_regular_Omega_epsi} we show the MSF in the plane $(\Omega,\epsilon_{1})$ for a fixed value of $\epsilon_2=0.01$ (panel (a)), and in the plane $(\Omega,\epsilon_{2})$ for a fixed value of $\epsilon_1=0.2$ (panel (b)). By using again $\hat{\epsilon}_1(\Omega)=\min \{\epsilon >0: \mathrm{MSF}(\epsilon,\epsilon_2,\Omega)<0\}$, for fixed $\epsilon_2$, and similarly $\hat{\epsilon}_2(\Omega)$, we can conclude that $\hat{\epsilon}_1(\Omega)<\hat{\epsilon}_1(0)\sim 1.4$ and $\hat{\epsilon}_2(\Omega)<\hat{\epsilon}_2(0)\sim 0.04$ and thus supporting again our claim that time-varying structures allow to achieve synchronization easier.

\section{Conclusions}
\label{sec:concl}
To sum up we have here introduced and studied a generalized framework for the emergence of global synchronization on time-varying higher-order networks and developed a theory for its stability without imposing strong restrictions on the functional time evolution of the higher-order structure. We have demonstrated that the latter can be examined by extending the Master Stability Function technique to the novel framework for specific cases based either on the inter-node coupling scheme or the topology of the higher-order structure. Our findings reveal that the behavior of the higher-order network is represented by a matrix that changes over time and possesses skew symmetry. This matrix is derived from the time-dependent evolution of the eigenvectors of the higher-order Laplacian. Additionally, the eigenvalues associated with these eigenvectors can also vary over time and have an impact on shaping the evolution of the introduced disturbance. We have validated the proposed theory on time-varying hypergraphs of coupled Stuart-Landau oscillators and chaotic Lorenz systems, and the results obtained indicate that incorporating temporal aspects into group interactions can facilitate synchronization in higher-order networks compared to static ones.
\par The framework and concepts presented in this study create opportunities for future research on the impact of temporality in systems where time-varying group interactions have been observed but not yet thoroughly explored due to the absence of a suitable mathematical setting. Importantly, the fact that our theory does not require any restrictions on the time evolution of the underline structure could offer the possibility to apply it for a diverse range of applications other than synchronization.

\bibliographystyle{apsrev4-1}
\bibliography{bib_temporal_hoi}

\onecolumngrid
\appendix 
\section{Non-invasive couplings}
\label{app:noninv}

Here we will discuss the results corresponding to a slightly more general hypothesis for $\vec{g}^{(d)}$, namely to be non-invasive, i.e.,
\begin{equation}
\label{eq:noninv}
 \vec{g}^{(d)}(\vec{s},\dots,\vec{s})=0\quad \forall d=1,\dots,D\, ,
\end{equation}
whose goal is again to guarantee that the coupling term in Eq.~\eqref{eq:dyn2} vanishes once evaluated on the orbit $(\vec{s}(t),\dots,\vec{s}(t))^\top$. Indeed by using again $\vec{x}_i=\vec{s}+\delta\vec{x}_i$ and expanding Eq.~\eqref{eq:dyn2} up to the first order we get
\begin{equation}
\label{eq:linearizedbis}
\delta\dot{\vec{x}}_i  =  \mathbf{J}_{f}\delta\vec{x}_i+\sum_{d=1}^D q_d\sum_{j_1,\dots,j_d=1}^n B_{ij_1\dots j_d}(t)  \left[  \frac{\partial \vec{g}^{(d)}}{\partial \vec{x}_{i}}\Big\rvert_{(\vec{s},\dots,\vec{s})}\delta\vec{x}_{i}+\frac{\partial \vec{g}^{(d)}}{\partial \vec{x}_{j_1}}\Big\rvert_{(\vec{s},\dots,\vec{s})}\delta\vec{x}_{j_1}+ \dots+  \frac{\partial \vec{g}^{(d)}}{\partial \vec{x}_{j_d}}\Big\rvert_{(\vec{s},\dots,\vec{s})}\delta\vec{x}_{j_d}\right]\, ;
\end{equation}
from Eq.~\eqref{eq:noninv} we can obtain
\begin{equation*}
\frac{\partial \vec{g}^{(d)}}{\partial \vec{x}_{i}}\Big\rvert_{(\vec{s},\dots,\vec{s})}+\frac{\partial \vec{g}^{(d)}}{\partial \vec{x}_{j_1}}\Big\rvert_{(\vec{s},\dots,\vec{s})}+ \dots+  \frac{\partial \vec{g}^{(d)}}{\partial \vec{x}_{j_d}}\Big\rvert_{(\vec{s},\dots,\vec{s})}=0\, ,
\end{equation*}
and thus rewrite~\eqref{eq:linearizedbis} as follows
\begin{equation}
\label{eq:linearizedtris}
\delta\dot{\vec{x}}_i  =  \mathbf{J}_{f}\delta\vec{x}_i+\sum_{d=1}^D q_d\sum_{j_1,\dots,j_d=1}^n B_{ij_1\dots j_d}(t)  \left[  \frac{\partial \vec{g}^{(d)}}{\partial \vec{x}_{j_1}}\Big\rvert_{(\vec{s},\dots,\vec{s})}(\delta\vec{x}_{j_1}-\delta\vec{x}_{i})+ \dots+  \frac{\partial \vec{g}^{(d)}}{\partial \vec{x}_{j_d}}\Big\rvert_{(\vec{s},\dots,\vec{s})}(\delta\vec{x}_{j_d}-\delta\vec{x}_{i})\right]\, .
\end{equation}
Recalling the definition of $k^{(d)}_{ij}$ given in Eq.~\eqref{eq:kij} we get
\begin{equation}
\label{eq:linearizedquat}
\delta\dot{\vec{x}}_i  =  \mathbf{J}_{f}\delta\vec{x}_i+\sum_{d=1}^D q_d (d-1)!\left[\sum_{j_1=1}^n k^{(d)}_{ij_1}(t) \frac{\partial \vec{g}^{(d)}}{\partial \vec{x}_{j_1}}\Big\rvert_{(\vec{s},\dots,\vec{s})}(\delta\vec{x}_{j_1}-\delta\vec{x}_{i})+ \dots+ \sum_{j_l=1}^n k^{(d)}_{ij_d}(t) \frac{\partial \vec{g}^{(d)}}{\partial \vec{x}_{j_d}}\Big\rvert_{(\vec{s},\dots,\vec{s})}(\delta\vec{x}_{j_d}-\delta\vec{x}_{i})\right]\, .
\end{equation}
By using the definition of the higher-order Laplace matrix~\eqref{eq:Lij} we eventually obtain
\begin{equation}
\label{eq:linearizedpent}
\delta\dot{\vec{x}}_i  =  \mathbf{J}_{f}\delta\vec{x}_i-\sum_{d=1}^D q_d\sum_{j=1}^n L^{(d)}_{ij}(t) \left[\frac{\partial \vec{g}^{(d)}}{\partial \vec{x}_{j_1}}\Big\rvert_{(\vec{s},\dots,\vec{s})}+ \dots+ \frac{\partial \vec{g}^{(d)}}{\partial \vec{x}_{j_d}}\Big\rvert_{(\vec{s},\dots,\vec{s})}\right]\delta\vec{x}_{j}\, .
\end{equation}

Let us consider now a particular case of non-invasive function, we assume thus there exists a function $\vec{\varphi}:\mathbb{R}^m\rightarrow \mathbb{R}^m$, such that $\vec{\varphi}(0)=0$ and define
\begin{equation}
\label{eq:specnoninv}
g^{(d)}(\vec{x}_i,\vec{x}_{j_1},\dots,\vec{x}_{j_d})=\sum_{\ell=1}^d\vec{\varphi}(\vec{x}_i-\vec{x}_{j_\ell})\, ,
\end{equation}
then
\begin{equation*}
 \frac{\partial \vec{g}^{(d)}}{\partial \vec{x}_{j_\ell}} = -\mathbf{J}_\varphi(0) \, ,
\end{equation*}
where $\mathbf{J}_\varphi(0)$ is the Jacobian of the function $\vec{\varphi}$ evaluated at $0$. In conclusion~\eqref{eq:linearizedpent} rewrites as follows
\begin{equation}
\label{eq:linearizedpent2}
\delta\dot{\vec{x}}_i  =  \mathbf{J}_{f}\delta\vec{x}_i-\sum_{d=1}^D q_d\sum_{j=1}^n L^{(d)}_{ij}(t) (-d)\mathbf{J}_\varphi(0) \delta\vec{x}_{j}=\mathbf{J}_{f}\delta\vec{x}_i+\sum_{j=1}^n G_{ij}(t)\mathbf{J}_\varphi(0) \delta\vec{x}_{j}\, ,
\end{equation}
where $\mathbf{G}(t)=\sum_{d=1}^Dd q_d\mathbf{L}^{(d)}(t)$ can be considered as an effective time-varying simplicial complex or hypergraph.
\par Let us now observe that the effective matrix $\mathbf{G}(t)$ is a Laplace matrix; it is non-positive definite (as each one of the $\mathbf{L}^{(d)}(t)$ does for any $d=1,\dots, D$ and any $t>0$), it admits $\mu^{(1)}=0$ as eigenvalue associated to the eigenvector $\phi^{(1)}=(1,\dots,1)^\top$ and it is symmetric. So there exist a orthonormal time-varying eigenbasis, $\phi^{(\alpha)}(t)$, $\alpha=1,\dots,n$, for $\mathbf{G}(t)$ with associated eigenvalues $\mu^{(\alpha)} \leq 0$. Similar to before, we define the $n\times n$ time dependent matrix $\mathbf{c}(t)$ that quantifies the projections of the time derivatives of the eigenvectors onto the independent eigendirections, namely
\begin{equation}
\label{eq:cab2}
\frac{d \vec{\phi}^{(\alpha)}}{dt}(t)=\sum_{\beta}c_{\alpha\beta}(t)\vec{\phi}^{(\beta)}(t)\quad\forall \alpha=1,\dots, n\, .
\end{equation}
By recalling the orthonormality condition $ \left(\vec{\phi}^{(\alpha)}(t)\right)^\top\cdot \vec{\phi}^{(\beta)}(t)=\delta_{\alpha \beta}$ we can again straightforwardly conclude that $\mathbf{c}$ is a real skew-symmetric matrix with a null first row and first column, i.e., $c_{\alpha\beta}+c_{\beta\alpha}=0$ and $c_{1\alpha}=0$.
\par Thereafter, we consider Eq.~\eqref{eq:linearizedpent2}, and we project it onto the eigendirections, namely we introduce $\delta\vec{x}_i=\sum_\alpha \delta\hat{\vec{x}}_{\alpha}\phi^{(\alpha)}_i$ and recalling the definition of $\mathbf{c}$ we obtain
\begin{equation}
\label{eq:GLHGlinalpha3_app}
\frac{d\delta\hat{\vec{x}}_{\beta}}{dt} = \sum_\alpha c_{\beta\alpha}(t)\delta\hat{\vec{x}}_{\alpha}+\left[\mathbf{J}_{f}+ \mu^{(\beta)}(t)\mathbf{J}_\varphi(0)\right]\delta\hat{\vec{x}}_{\beta}\, . 
\end{equation}
This is the required Master Stability Equation, solving which for the calculation of maximum Lyapunov exponents provide the condition for stability of the synchronous solution. 

\subsection{Synchronization of Stuart-Landau oscillators with non-invasive coupling assumption}
\label{ssec:noninvass}

\begin{figure}
	\centerline{
		\includegraphics[scale=0.33]{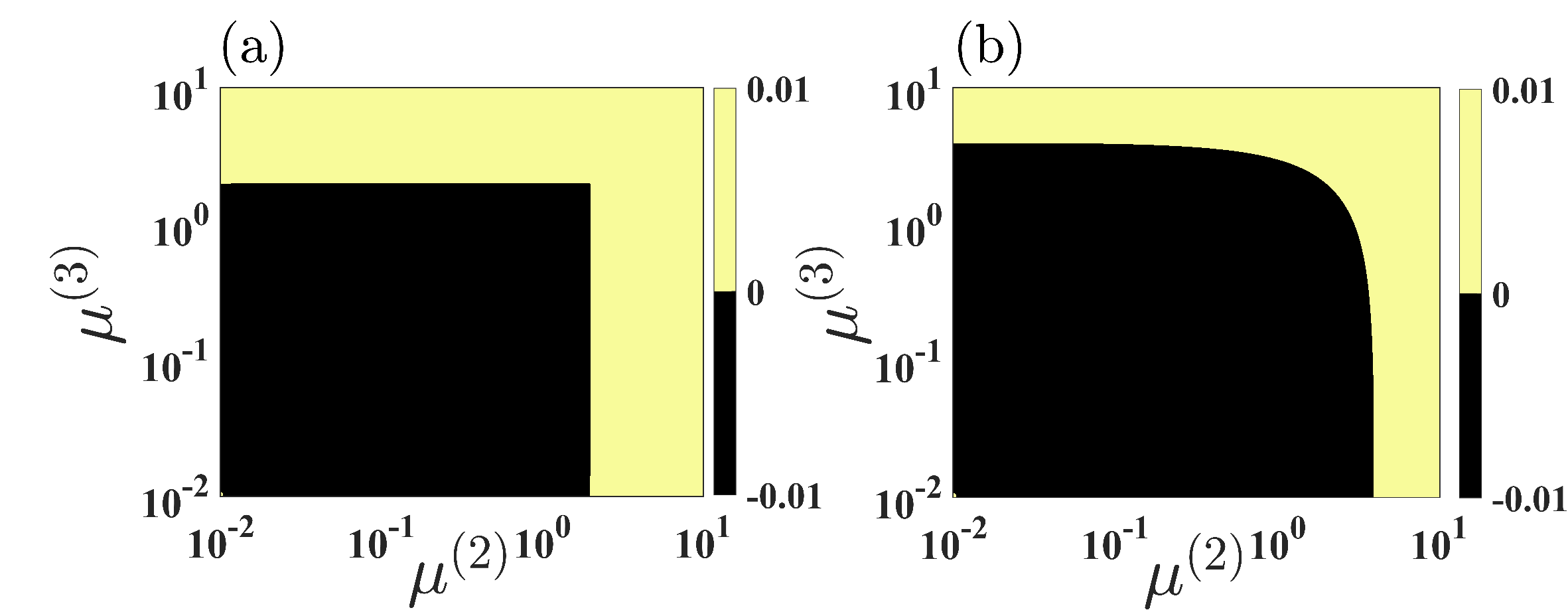}}
	\caption{{\bf Synchronization on time-varying higher-order network of coupled SL oscillators with non-invasive coupling configuration.} Region of synchrony and desynchrony are depicted by simultaneously varying $\mu^{(2)}$ and $\mu^{(3)}$ for two different values of $\Omega$ (a) $\Omega=0$, (b) $\Omega=2$, where the domain in black indicates the area of the stable synchronous solution. The range of the axes is characterized by considering the absolute values of the eigenvalues. All the other values are kept fixed at $\sigma=1.0+4.3i$, $\beta=1.0+1.1i$, $\varphi^\prime(0)=1$.}
	\label{SL_non_invasive_ev2_ev3}
\end{figure}

To validate the above results we again consider the SL oscillator with a particular case of non-invasive coupling function, namely we assume to exist a real function $\varphi$ such that $\varphi(0)=0$, $\varphi^\prime(0)\neq 0$ and
\begin{equation}
\label{eq:noninvvarphi}
\begin{array}{l}
g^{(1)}(w_1,w_2)=\varphi(w_1-w_2)\, , \text{ and } \\ g^{(2)}(w_1,w_2,w_3)=\varphi(w_1-w_2)+\varphi(w_1-w_3)\, .
\end{array}
\end{equation}
By reasoning as before, we get
\begin{equation}
\label{eq:linnoninv}
\begin{array}{l}
 \dfrac{d}{dt}\left(\begin{matrix} {\rho_j} \\{\theta_j}\end{matrix}\right) = 
\left(\begin{matrix}
 -2\sigma_\Re & 0\\-2\beta_\Im \frac{\sigma_\Re}{\beta_\Re} & 0
\end{matrix}\right)\left(\begin{matrix} {\rho_j} \\{\theta_j}\end{matrix}\right)+ \varphi^\prime(0)\sum_\ell  \left(q_1 L^{(1)}_{j\ell} +q_2 L^{(2)}_{j\ell}\right) \left(\begin{matrix}
1 & 0\\ 0 & -1
\end{matrix}\right)\left(\begin{matrix} {\rho_l} \\{\theta_l}\end{matrix}\right) .
\end{array}
\end{equation}
By using again the eigenvectors $\phi^{(\alpha)}(t)$, eigenvalues $\mu^{(\alpha)}(t)$ of $\mathbf{G}(t)$ and the matrix $\mathbf{c}$ (see Eq.~\eqref{eq:cab2}), we can rewrite the previous formula as
\begin{equation}
\label{eq:linnoninv2}
\begin{array}{l}
 \dfrac{d}{dt}\left(\begin{matrix} {\rho_\beta} \\{\theta_\beta}\end{matrix}\right) = \sum_\alpha c_{\beta\alpha}\left(\begin{matrix} {\rho_\alpha} \\{\theta_\alpha}\end{matrix}\right)+\biggl[\left(\begin{matrix}
 -2\sigma_\Re & 0\\-2\beta_\Im \frac{\sigma_\Re}{\beta_\Re} & 0
\end{matrix}\right)  + \varphi^\prime(0)\mu^{(\beta)} \left(\begin{matrix}
1 & 0\\ 0 & -1
\end{matrix}\right)\biggr]\left(\begin{matrix} {\rho_\beta} \\{\theta_\beta}\end{matrix}\right).
\end{array}
\end{equation}
Figure \ref{SL_non_invasive_ev2_ev3} represent the result for the non-invasive coupling assumption. Here, we consider the non-invasive function so that $\varphi^\prime(0)=1$ and the skew-symmetric projection matrix $\mathbf{c}$ is considered constant throughout the analysis as earlier. Here we show the level sets of the MSF as a function of the eigenvalues $\mu^{(2)}$ and $\mu^{(3)}$ while keeping the remaining parameters in Eq.~\eqref{eq:linnoninv2} fixed at generic nominal values. In panel (a), we consider a static hypergraph, i.e., $\Omega=0$, while in the (b) panel, a time-varying hypergraph, i.e., $\Omega=2$, negative values of MSF are reported in black, and they correspond thus to a global synchronous state, positive values of MSF are shown in yellow; one can clearly appreciate that in the case of the time-varying hypergraph, the MSF is negative for a much larger set of eigenvalues $\mu^{(2)}$ and $\mu^{(3)}$ and thus the SL system can achieve synchronization more easily.

\section{Structure of the small hypergraph}
\label{sec:hypergraph}
The goal of this section is to provide more details about the construction of the simple time-varying hypergraph used as support for the numerical simulations in the main text. To start with we need to obtain the time-evolution of eigenvectors $\vec{\psi}^{(\alpha)}(t)$, which follows the equation 
\begin{equation}
    \begin{array}{l}
       \dfrac{d\vec{\psi}^{(\alpha)}}{dt}=\sum\limits_{\alpha}b_{\beta\alpha}\vec{\psi}^{(\alpha)}\, ,
    \end{array}
\end{equation}
where the matrix $\mathbf{b}$ has been given in Eq.~\eqref{eq:bmatrixOmega}. The eigenvector associated with the least eigenvalue $\Lambda^{(1)}=0$ is constant and is given by $\vec{\psi}^{(1)}=\frac{1}{\sqrt{3}}(1,1,1)^{\top}$. The other two eigenvectors are obtained by solving the previous equation and are represented as $\vec{\psi}^{(2)}(t)=\vec{v}_{1}\cos(\Omega t)+\vec{v}_{2}\sin(\Omega t)$ and
$\vec{\psi}^{(3)}(t)=-\vec{v}_{1}\sin(\Omega t)+\vec{v}_{2}\cos(\Omega t)$, where $\vec{v}_{1}$, $\vec{v}_{2}$ are the unknown vectors that should be determined using the constraints to have orthonormal eigenbasis for every $t$.
Following a few steps of calculation, we can obtain the other two eigenvectors as follows
\begin{equation}\label{eigenvectors}
	\begin{array}{l}
		\vec{\psi}^{(2)}(t)=\dfrac{1}{\sqrt{6}}\begin{pmatrix}
			1 \\ -2 \\ 1
		\end{pmatrix}\cos(\Omega t)+\dfrac{1}{\sqrt{2}}\begin{pmatrix}
		-1 \\ 0 \\ 1
	\end{pmatrix}\sin(\Omega t)  \;\; \mbox{and}, \\
    \vec{\psi}^{(3)}(t)=-\dfrac{1}{\sqrt{6}}\begin{pmatrix}
    	1 \\ -2 \\ 1
    \end{pmatrix}\sin(\Omega t)+\dfrac{1}{\sqrt{2}}\begin{pmatrix}
    	-1 \\ 0 \\ 1
    \end{pmatrix}\cos(\Omega t). 
	\end{array}
\end{equation} 
Now recalling our assumption about constant eigenvalues and using the relation $\mathbf{L}^{(1)}_{ij}(t)=\sum\limits_{\alpha}\Lambda^{(\alpha)}\vec{\psi}^{(\alpha)}_{i}(t)\vec{\psi}^{(\alpha)}_{j}(t)$, we can obtain the entries of the pairwise Laplace matrix as
\begin{equation} \label{pairwise_lap}
	\begin{array}{l}
		{L}^{(1)}_{ij}(t)=\Lambda^{(2)}\vec{\psi}^{(2)}_{i}(t)\vec{\psi}^{(2)}_{j}(t)+\Lambda^{(3)}\vec{\psi}^{(3)}_{i}(t)\vec{\psi}^{(3)}_{j}(t),
	\end{array}
\end{equation}
where we use the fact that $\Lambda^{(1)}=0$ for all time $t$. Finally by using the relation between pairwise adjacency and Laplace matrices $L^{(1)}_{ij}(t)=A^{(1)}_{ij}(t)$, for $i\ne j$, we obtain the temporal evolution of the links as 
\begin{equation}\label{edge_evolution}
	\begin{array}{l}
		A^{(1)}_{12}(t)=\dfrac{1}{2}-\dfrac{1}{3}\cos(\frac{\pi}{3}+2\Omega t), \\\\
		A^{(1)}_{13} (t)= \dfrac{1}{2}+\dfrac{1}{3}\cos(2\Omega t), \\\\
		A^{(1)}_{23}(t)=\dfrac{1}{2}-\dfrac{1}{3}\cos(\frac{\pi}{3}-2\Omega t),
	\end{array}
\end{equation} 
where we have used the fact that the non-zero eigenvalues are given by $\Lambda^{(2)}=-1$ and $\Lambda^{(3)}=-2$. 
\par
\begin{figure}
	\centerline{
		\includegraphics[scale=0.33]{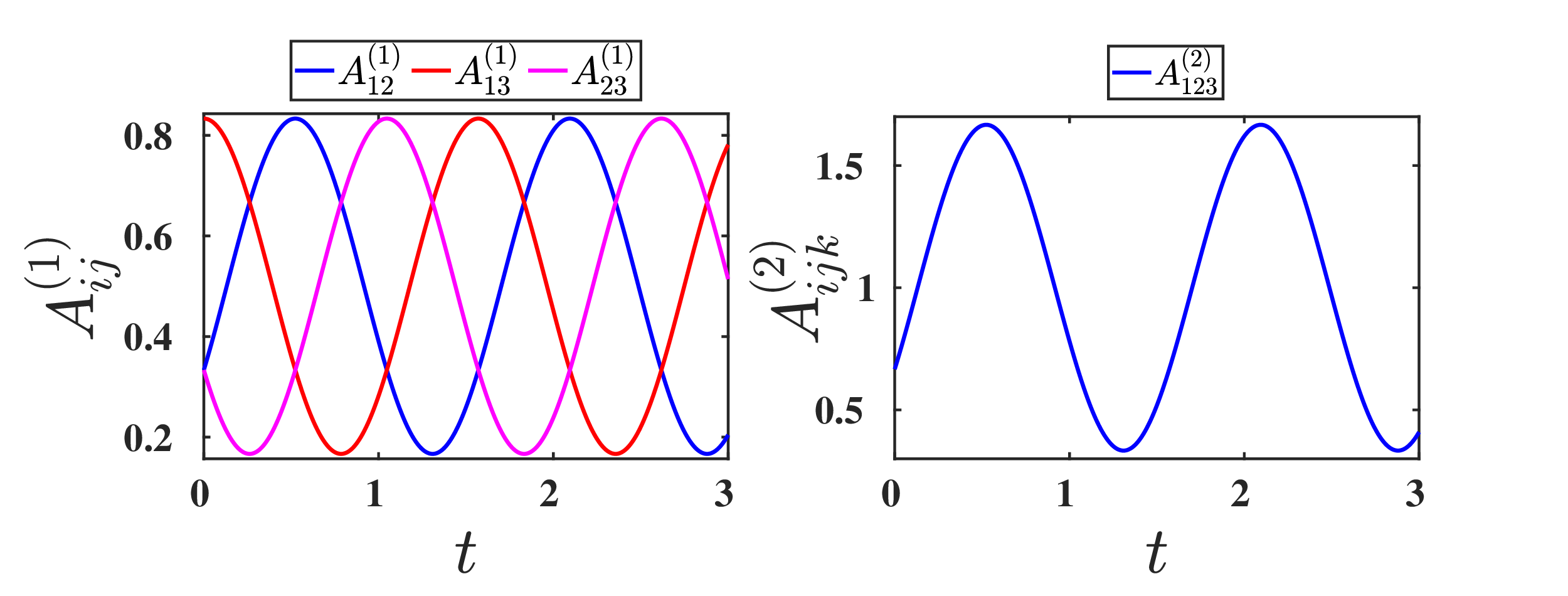}}
	\caption{{\bf Temporal evolution of edges and $3$-hyperedge obtained from Eqs.~\eqref{edge_evolution} and~\eqref{hyperedge_evolution} for a particular value of $\Omega=2$.}}
	\label{temporal evolution}
\end{figure}

 Again from the regular structure of the hypergraph, we have $\mathbf{L}^{(2)}(t)=\alpha_{2}\mathbf{L}^{(1)}(t)$, for all $t$. Therefore, following the relation~\eqref{pairwise_lap}, entries of the $2nd$-order Laplacian $\mathbf{L}^{(2)}$ can be represented as,   
\begin{equation} \label{hoi_lap}
	\begin{array}{l}
		L^{(2)}_{ij}(t)=\alpha_{2}[\Lambda^{(2)}\vec{\psi}^{(2)}_{i}(t)\vec{\psi}^{(2)}_{j}(t)+\Lambda^{(3)}\vec{\psi}^{(3)}_{i}(t)\vec{\psi}^{(3)}_{j}(t)].
	\end{array}
\end{equation}
Now, the definition of higher-order Laplacian implies that, $L^{(2)}_{ij}(t)=\sum\limits_{k}A^{(2)}_{ijk}(t)$, $i \ne j$. Hence, using the above relation and Eq. \eqref{hoi_lap}, we can obtain the temporal evolution of the $3$-hyperedge as
\begin{equation}\label{hyperedge_evolution}
	\begin{array}{l}
		A^{(2)}_{123}(t)=1-\dfrac{2}{3}\cos(\frac{\pi}{3}+2\Omega t),
	\end{array}
\end{equation}
where we have again used the fact that the non-zero eigenvalues are $\Lambda^{(2)}=-1$, and $\Lambda^{(3)}=-2$, and the value of the parameter $\alpha_{2}$ has been set $\alpha_{2}=2$. Due to the assumption of undirected hypergraph, we also trivially have, $A^{(2)}_{123}(t)=A^{(2)}_{\pi{(123)}}(t)$, where $\pi{(123)}$ indicates any permutation of $(123)$. Fig.~\ref{temporal evolution} portrays the temporal evolution of the links and $3$-hyperedge weights. To better understand the evolution of the hypergraph, we provide the graphical evolution of the hypergraph in the accompanying Supplementary Movie, together with the time evolution of the weights of the links $A^{(1)}_{ij}(t)$ and of the hyperedge $A^{(2)}_{123}(t)$.

\end{document}